\documentclass[12pt]{iopart}

\usepackage{iopams}
\usepackage{cite}
\expandafter\let\csname equation*\endcsname\relax
\expandafter\let\csname endequation*\endcsname\relax
\usepackage[T1]{fontenc}

\usepackage{amsmath}
\usepackage{commath}
\usepackage{iopams}
\usepackage{graphicx}
\usepackage[breaklinks=true,colorlinks=true,linkcolor=blue,urlcolor=blue,citecolor=blue]{hyperref}
\usepackage{color}
\usepackage{mathrsfs}

\def\P{\mathbb{P}}

\def\erf{\mathrm{erf}}
\def\erfc{\mathrm{erfc}}

\begin{document}

\title[]{One-dimensional lattice random walks in a Gaussian random potential}

\author{Silvio Kalaj$^{1,2}$, Enzo Marinari$^{1,3}$, Gleb Oshanin$^{2}$ \& Luca Peliti$^4$}
\address{$^1$ Dipartimento di Fisica, Sapienza Universit\`{a} di Roma, P.le A. Moro 2, I-00185, Roma, Italy}
\address{$^2$ Sorbonne Universit\'e, CNRS, Laboratoire de Physique Th\'eorique de la Mati\`{e}re Condens\'ee (UMR 7600), 4 Place Jussieu, 75252, Paris, Cedex 05, France}
\address{$^3$ INFN, Sezione di Roma 1 and Nanotech-CNR, UOS di Roma, P.le A. Moro 2, I-00185, Roma, Italy}
\address{$^4$ Santa Marinella Research Institute, I-00058, Santa Marinella, Italy}

\begin{abstract}
We study random walks evolving in continuous time on a one-dimensional lattice where each site $x$ hosts a quenched random potential $U_x$. The potentials on different sites are independent, identically distributed  Gaussian random variables.  
We analyze three distinct models that specify how the transition rates depend on 
$U_x$: the random-force-like model, random walks with randomized stepping times, and the Gaussian trap model. Our analysis focuses on five key disorder-dependent quantities defined for a finite chain with 
$N$ sites: the probability current, its reciprocal (the resistance), the splitting probability  $E_-$, the mean first-passage time $T_N$, and the diffusion coefficient $D_N$ in a periodic chain. By determining the moments of these random variables, we demonstrate that the probability current and resistance are not self-averaging, which leads to pronounced differences between their average and typical behaviors. In contrast, $E_-$, $T_N$ and $D_N$ become self-averaging when $N \to \infty$, though they exhibit strong sample-to-sample fluctuations for finite $N$.

\end{abstract}

Keywords:  Random walks in disordered media, Gaussian random potential, self-averaging vs non-self-averaging

\vspace{0.2in}

\today

submitted to \JPA

\maketitle

\section{Introduction}     

Understanding the various aspects of a particle's random motion in disordered environments presents a significant challenge, stemming from both mathematical and physical complexities, as well as a wide range of practical applications. In many real-world systems, the medium through which a particle moves is highly irregular, influenced by defects, impurities, fluctuations, and other factors. These irregularities create a random environment that affects local transition rates and often alters the behavior of averaged transport characteristics.
Since the early studies that introduced several foundational results, along with various analytical approaches (see, for example, \cite{as,der,luck,derdiff,fis,durrett1,durrett2,zwan,touya,dean3}), substantial progress has been made across different areas of this vast domain. Notable topics include diffusion in random forcing landscapes \cite{sinai,pom,enzo1,enzo2,fish,kesten,pel1,deandean,d0,d1,d2,d3}, in the presence of quenched random traps (see, e.g., \cite{jpb,leticia,mont,bertin,bertin2,ralf10,ralf11,ralf12,aki1,aki2}), random stratified flows \cite{mdm,blu,kay}, and flows on various types of random lattices \cite{man1,man2}.
The impact of disorder -- such as sequence heterogeneity -- on random transport in biological systems has been studied in various contexts. These include polynucleotide translocation through nanopores, the movement of molecular motors or proteins along DNA \cite{r1,r2,r3,r4}, the dynamics of twist-induced denaturation bubbles in long DNA strands \cite{22}, and the behavior of the boundary between coil and ordered helix phases during heteropolymer melting \cite{pdg} (see also \cite{d11}). A rich body of results and concepts has been summarized in several comprehensive reviews \cite{al,pel,kehr,bou,hughes,dean,bogachev}. Despite this progress, the problem remains a vibrant area of research, with many aspects still not fully understood and being the subject of ongoing studies. For example, recent papers have examined the influence of correlated disorder on transport properties \cite{baruch}, the dynamics of active particles such as bacteria in disordered environments \cite{olson,licata,zeitz,sant,mat}, and sequence-reading diffusion model that mimics in some effective way the random translocation of knots along chemically heterogeneous DNA strands \cite{kalaj}.

In this paper, we revisit the long-standing problem of random walks in the presence of a quenched random Gaussian potential (Gaussian white noise), a topic that has received a significant attention in the past (see, e.g., \cite{dean}). Specifically, we study dynamics of a particle that undergoes, in continuous time, a random walk between nearest-neighboring sites of a one-dimensional lattice, where each site $x$ hosts a quenched random potential $U_x$, which influences the local transition rates. The variables $U_x$ are independent and identically distributed (iid) random variables, with a common Gaussian distribution. 
This model can be also viewed from a somewhat different perspective. Namely, it can be considered  as an effective description, in terms of the Fick-Jacob's approximation (see, e.g., \cite{1,2}),   of diffusion in a narrow two- or three-dimensional
 channel with Gaussian-distributed  
 random wall corrugation. 

To define the transition rates as functions of the on-site potentials, we consider three distinct scenarios: 
 a) the "random force"-like model \cite{bou}, b) a continuous-time random walk \cite{mw} with an exponential waiting-time distribution and  the transition probabilities dependent on the values of potentials at the neighboring sites, and c) a Gaussian random trap model (see, e.g., \cite{jpb,leticia,mont,bertin,bertin2,ralf10,ralf11,ralf12,aki1,aki2,ralf13} and \cite{bou}). We will provide a more detailed definition of the dynamics for all three models in the following sections.

  Our analysis focuses on five \textit{realization-dependent} random variables :\\ a) The steady-state probability current through a finite chain consisting of $N$ sites. \\ b) Its reciprocal value, which represents the resistance of a finite interval with $N$ sites with respect to a random transport. \\ c) The splitting probability - the probability that the particle starting at the site $x_0$ in a finite interval $(0,N)$ reaches the left extremity of the interval first, without ever visiting the right end of the interval.\\
d) The mean first-passage time through a finite interval of length $N$. \\
e) The diffusion coefficient in a finite periodic chain with $N$ sites.\\
While the statistics of the last two quantities have been explored in previous studies and some well-established results have been  obtained (see, e. g., \cite{bou,kehr,dean}), a comprehensive analysis of the first three variables is lacking at present. It is also important to note that all the random variables under study are formally defined in the limit $t \to \infty$. Consequently, our analysis does not address transient effects, such as aging and the ensuing sub-diffusive behavior, which have been extensively discussed in the literature on the random force and random traps models (see, e. g., \cite{bou}).

Our objective is to calculate the leading asymptotic forms of the moments of these random variables in the large-$N$ limit, as well as to understand how they depend on the strength of disorder. For intermediate values of $N$, we will conduct numerical analyses, employing exact numerical averaging of the formal expressions that define these properties for a given sequence of $U_x$ values. Ultimately, we aim to address a  conceptually important question whether these transport properties exhibit self-averaging with respect to different realizations of disorder in the limit $N \to \infty$. In quest  for the answer, we will examine the large-$N$ asymptotic forms of the key parameter - the relative variance
\begin{align}
\label{R}
R_{\zeta} = \frac{\langle \zeta^2 \rangle - \langle \zeta \rangle^2}{\langle \zeta \rangle^2}  \,,
\end{align} 
where the angle brackets here and henceforth denote averaging with respect to realizations of disorder, while  $\zeta$ is one of the above-mentioned random variables of interest. 
This parameter establishes a standard criterion (see, for example, \cite{pastur}) that allows us to conclude whether the realization-dependent random variable $\zeta$ is self-averaging ($R_{\zeta} \to 0$ as $N \to \infty$) or not ($R_{\zeta} \to {\rm const}$ as $N \to \infty$). Determining which of the two possible outcomes is realized is of a considerable conceptual importance: in a self-averaging system, a single sufficiently large sample can provide a representative insight into the behavior of the entire statistical ensemble. Conversely, in the absence of self-averaging, the value of an observable derived from a single, even large, sample will not be representative and must be assessed from an ensemble of samples. 
Consequently, one may expect a much more significant scatter of the data for a single realization of disorder 
 in the latter than in the former case.
It is also important to note that for finite $N$, the relative variance $R_{\zeta}$ is always greater than zero, indicating that fluctuations are significant for finite samples.
Note as well that the limit $N \to \infty$  itself is very meaningful for the diffusion coefficient, as it reflects the behavior expected in infinite systems. In contrast, for other properties of interest, this limit will only highlight a trend in the behavior of fluctuations. Specifically, for quantities such as currents, resistances, splitting probabilities, and the mean first-passage times, the size $N$ remains fundamentally finite.

The paper is outlined as follows: In Sec. \ref{sec:2} we present our model 
and describe three different dynamical scenarios.  In Sec. \ref{sec:3}
we concentrate on the statistical properties of the realization-dependent currents through a finite chain and of its reciptocal value - the resistance of a finite chain with respect to a random transport. In Sec. \ref{sec:4} we discuss the behavior of the realization-dependent splitting probabilities. Section \ref{sec:5} is devoted to the analysis of the mean first-passage times through finite intervals. Next, in Sec. \ref{sec:6} we evaluate the moments of the diffusion coefficient in period chains in the limit $N \to \infty$ for all three dynamical scenarios. In Sec. \ref{sec:7} we conclude with a brief recapitulation of our results. Some derivations are relegated to Appendices. In \ref{A} we present the derivation of the probability currents through a finite chain with fixed disorder for all three dynamical scenarios, while in \ref{B} we express the realization-dependent 
splitting probability through the 
resistances of the intervals from the right and from the left of the starting point $x=x_0$.  Lastly, in \ref{C}, capitalizing on the exact result due to Derrida \cite{derdiff}, we derive exact compact expressions for the realization-dependent  
diffusion coefficient on a periodic chain for all three dynamical scenarios.

\section{Model and dynamical scenarios}
\label{sec:2}

Consider a one-dimensional regular lattice with inter-site spacing $a$. Without any lack of generality,  we set $a=1$ in what follows. We assign to each site $x$ ($x$ is an integer)
of this lattice a quenched random variable $U_x$ and assume that all
$U_x$-s are independent, identically distributed (iid) random variables with the common 
Gaussian distribution. When $U_x$ are defined on the entire real line,  $-\infty < U_x < \infty$, we have
 \begin{equation}
 \label{Gauss}
P(U) = \frac{1}{\sqrt{2 \pi \sigma^2}} \exp\left(- \frac{U^2}{2 \sigma^2}\right) \, ,
\end{equation}
while if $U_x$ may attain only negative real values\footnote{See below the definition of the Gaussian random trap model.},  $-\infty < U_x \leq 0$, the probability density function of $U_x$ obeys 
 \begin{equation}
 \label{Gaussm}
P_-(U) = \sqrt{\frac{2}{ \pi \sigma^2}} \exp\left(- \frac{U^2}{2 \sigma^2}\right) \, .
\end{equation}
In what follows we will also use
 auxiliary variables 
$\phi_x$, defined as
\begin{align}
\label{phi}
\phi_x = \exp\Big(\frac{\beta U_x}{2}\Big) \,,
\end{align}
which will permit us to somewhat ease the notations and also to write down all the properties under study, which are functionals  of a set of all $\phi_x$-s, in a very compact and transparent way. 
For Gaussian iid variables $U_x$, the variables
$\phi_x$ are also iid random variables with the common log-normal probability density function. 
For $U_x$ defined on the entire axis
\begin{equation}
\label{Pphi}
P(\phi) = \sqrt{\frac{2}{\pi \beta^2 \sigma^2}} \frac{1}{\phi} \exp\left(- \frac{2}{\beta^2 \sigma^2} \ln^2(\phi)\right) \,, \quad 0 \leq \phi < \infty \,,
\end{equation}
while for $-\infty < U_x \leq 0$ we have
\begin{equation}
\label{Pphim}
P_-(\phi) = \sqrt{\frac{8}{\pi \beta^2 \sigma^2}} \frac{1}{\phi} \exp\left(- \frac{2}{\beta^2 \sigma^2} \ln^2(\phi)\right) \,, \quad 0 \leq \phi \leq 1 \,.
\end{equation}
In both cases the variables $\phi_x$ have finite moments of \textit{arbitrary} positive or negative (not necessarily integer) order $q$.

The random walk of a particle in such a random frozen environment 
is defined using three different scenarios: 
 
\subsection{Scenario  I:    Random force-like model}

Here we follow the rules of the so-called "random force" model (see, e.g., \cite{bou} for a review) defining on 
each link that connects sites $x$ and $x+1$ a random force $F_{x,x+1} = -(U_{x+1} - U_x)/a$, ($a = 1$).  Note that this force varies randomly from site to site, but is not statistically \textit{independent} from the forces defined on two neighboring sites, as it is the case in, e.g., the Temkin or Sinai models \cite{sinai,pom,enzo1,enzo2,fish,kesten,pel1,deandean,d0,d1,d2,d3}. In our setting the random force 
is the difference of two on-site potentials, which are the iid random variables\footnote{To prevent any confusion, we refer therefore to this dynamic scenario as the "random-force-like" model, since the term "random-force" model is typically reserved for situations in which the forces themselves are independent and identically distributed random variables.}.
We stipulate then that the particle performs activated hopping motion in a continuous-time between the nearest-neighboring sites with the transition rates $W_{x,x+1}$ and $W_{x,x-1}$ from the host site $x$ to the target sites $x+1$ or $x-1$, respectively, which obey
\begin{align}
\begin{split}
\label{rates}
&W_{x,x\pm1} = W_0 \exp\left(\frac{\beta}{2} \left(U_x - U_{x\pm1}\right)\right)= W_0 \frac{\phi_x}{\phi_{x\pm1}} \,,\\ 
&W_{x\pm1,x} = W_0 \exp\left(\frac{\beta}{2} \left(U_{x\pm1} - U_{x}\right)\right) = W_0 \frac{\phi_{x\pm1}}{\phi_x} \,,
\end{split}
\end{align} 
where $\beta$ is the reciprocal temperature measured in units of the Boltzmann constant, while $W_0$ is the characteristic rate which defines the frequency of jumps in absence of disorder.  In this scenario, the probability $P_x(t)$ of finding the particle at site $x$ at time instant $t$ obeys the master equation
\begin{align}
\label{me}
\dot{P}_x(t) = W_{x+1,x} P_{x+1}(t)  + W_{x-1,x} P_{x-1}(t) - \left(W_{x,x+1} + W_{x,x+1} \right) P_x(t) \,,
\end{align} 
where the dot denotes the time derivative. Such a model evidently 
satisfies the detailed balance and, on a finite periodic chain with $N$ sites, $x = 1, 2, \ldots, N$,
the normalized to unity position probability $P_x$ evolves towards an equilibrium state with 
\begin{equation}
\label{dens1}
P_x(t = \infty) = \frac{\exp( - \beta U_x)}{\sum_{y=1}^N \exp( - \beta U_y)} = \frac{\phi_x^{-2}}{\sum_{y=1}^N \phi_y^{-2}} \,.
\end{equation} 

\subsection{Scenario  II: Random walk with randomized stepping-times}

We aim now to work out a somewhat different dynamical scenario which yields a continuous-time master equation with
the normalized transition \textit{probabilities}, rather than the rates.  To this end, consider first an auxiliary random walk problem 
in which a particle steps at each tick $n = 1, 2, 3, \ldots $ of the clock  between the nearest-neighboring sites of a lattice of integers $x$. The probability $P_x(n)$ of finding the particle at site $x$ at discrete time moment $n$ is governed by a general equation for a random walk \cite{hughes} :
\begin{equation}
\label{disc}
P_x(n) = p_{x+1,x} P_{x+1}(n-1) + p_{x-1,x} P_{x-1}(n-1) \,, 
\end{equation}
where $p_{x+1,x}$ and $p_{x-1,x}$ are (time-independent) probabilities that a particle hops from the site $x+1$ or from the site $x-1$, respectively, to the site $x$. The probabilities of hops from the site $x$ to either of the neighboring sites are normalized, $p_{x,x+1} + p_{x,x-1} = 1$, such that the particle cannot pause at $x$.   

Recalling that each site of the lattice hosts a random potential $U_x$, 
we define the transition 
\textit{probabilities} $p_{x,x\pm1}$ in the way which resembles 
the definition of the transition rates in the above Model I, except 
that these probabilities are normalized to unity. 
 That being, we stipulate that
\begin{equation}
\begin{split}
\label{3}
p_{x,x+1} &= Z^{-1}_x   \exp\left(\frac{\beta}{2}\Big[U_{x} - U_{x+1}  \Big]\right) \,, \\
p_{x,x-1} &=  1 - p_{x,x+1} = Z^{-1}_x \exp\left(\frac{\beta}{2} \Big[U_{x} - U_{x-1} \Big] \right) \,, 
\end{split}
\end{equation}
 where $Z_x$ is the normalization, 
 \begin{align}
 Z_x =  \exp\left(\frac{\beta}{2}\Big[U_{x} - U_{x+1}  \Big]\right)  +  \exp\left(\frac{\beta}{2} \Big[U_{x} - U_{x-1} \Big] \right)\,.
 \end{align}
The transition probabilities $p_{x,x+1}$ and $p_{x,x-1}$ in Eqs. \eqref{3} can be also conveniently expressed through auxiliary variables $\phi_x$ in Eq. \eqref{phi} to give
\begin{equation}
\begin{split}
\label{33}
p_{x,x+1} =  \frac{\phi_{x-1}}{\phi_{x-1} + \phi_{x+1}}\,, \quad
p_{x,x-1} =  \frac{\phi_{x+1}}{\phi_{x-1} + \phi_{x+1}} \,. 
\end{split}
\end{equation}
Note that the definition of the transition probabilities in Eq. \eqref{3} is essentially the same as in the "random walk on a random hillside" model \cite{durrett1,durrett2}, except that in the latter the potential of the target site is taken in the middle of the link connecting the host and the target sites. In our model 
the transition probabilities
 involve the values of potentials on the host and the target sites themselves.  
 
Before we proceed, it might be also instructive  to formally rewrite  eqs. \eqref{3} as
\begin{equation}
\begin{split}
\label{30}
p_{x,x+1} = \frac{1}{1  + \exp\left( - \beta F_{x-1,x+1}\right)} \,, \quad
p_{x,x-1} = \frac{1}{1 +  \exp\left(\beta F_{x-1,x+1} \right)} \,, 
\end{split}
\end{equation}
where $F_{x-1,x+1} = - (U_{x+1} - U_{x-1})/2$ can be identified as the force acting on the particle being at site
 $x$. In this way, the particle being at site $x$ "feels" only the values of potentials at two adjacent sites and  
   the probability to jump towards the neighboring site with a lower energy is bigger than the probability to make a jump in the opposite direction, regardless of the actual value of $U_x$.  

We turn to a \textit{randomized stepping-time} version of the above model, which will be our Model II. We  
suppose now that the random walk evolves in continuous-time and  in each realization of the process the particle steps from the site it occupies at random 
time instants $t_1$, $t_2$, $t_3, \ldots$, where the time-intervals  $\triangle_1 = t_1 - t_0$, $(t_0 =0)$, $\triangle_2 = t_2 - t_1$, $\triangle_3 = t_3 - t_2$, $\ldots$, are iid random variables with the probability density function
\begin{equation}
\label{mem}
P(\triangle) = \exp\left(- \frac{\triangle}{\delta t}\right)/\delta t \,.
\end{equation}  
The sequence $t_1$, $t_2$, $t_3, \ldots$ is therefore a Poisson point process with density $P(\triangle)$ and the mean value of the time-interval between the successive steps is equal to $\delta t$, such that  
the ensuing hopping process is simply
the 
Montroll-Weiss continuous-time random walk (CTRW) \cite{mw} (see also \cite{hughes} for an overview) 
in which the 
waiting times are iid random variables 
with an exponential distribution.

To define the time-evolution of the position probability $P_x = P_x(t)$ for Model II, 
we 
take advantage of the analysis in \cite{katja} (see also \cite{hughes}), which showed that 
for any value of time $t$ 
the time-evolution of the latter probability for
the above-defined continuous-time random walk is governed, in place of the discrete-time Eq. \eqref{disc}, 
 by the master equation of the form : 
\begin{align}
\label{ME}
\delta t \dot{P}_x(t) =  p_{x-1,x} P_{x-1}(t) + p_{x+1,x} P_{x+1}(t) -  P_x(t) \,,
\end{align}
where the transition probabilities are defined in eqs. \eqref{3} (or eqs. \eqref{30}).
The master equation \eqref{ME} is Markovian, as the one for the Model I, 
so that the future evolution is sensitive only to the present state while the detailed history is irrelevant.
 At first glance, this may seem somewhat surprising because in the continuous-time random walk  
 the particle steps at random time instants and therefore, there is a memory (albeit a short-ranged) on 
 the time instant 
 when the previous stepping event took place. As shown in \cite{katja}, a Poisson point process with the density $P(\triangle)$  in Eq.  \eqref{mem} is a \textit{unique} example of a stochastic stepping process 
 for which the Markovian master equation \eqref{ME}, corresponding to the process in Eq. \eqref{disc} with randomized stepping times, is exact.

 Note, as well, that Eq. \eqref{ME} evidently satisfies the detailed balance 
 and its stationary solution on a periodic chain with $N$ sites  (see \cite{sasha} for the definition 
 of $P_x(t=\infty)$ in terms of the transition probabilities or rates) has a bit more complicated form than the one in Eq. \eqref{dens1}. For any given sequence of $U_x$ it reads
 \begin{align}
 \begin{split}
  \label{dens2}
 P_x(t = \infty) &= \frac{1}{2} \left(\exp\left(-\frac{\beta}{2} (U_{x-1} + U_x)\right) + \exp\left(- \frac{\beta}{2} (U_x + U_{x+1})\right) \right) \\ &\times \left(\sum_{y = 1}^N \exp\left(- \frac{\beta}{2} (U_y + U_{y+1})\right)\right)^{-1} \\
 &= \frac{1}{2} \left(\frac{1}{\phi_{x-1} \phi_x} + \frac{1}{\phi_x \phi_{x+1}}\right)  \left(\sum_{y = 1}^N \frac{1}{\phi_x \phi_{x+1}}\right)^{-1} \,, 
\end{split}
\end{align} 
with $U_{N+1} = U_1$ and $\phi_{N+1} = \phi_1$.

We finally remark that although the master equations \eqref{me} and \eqref{ME} have quite different forms, 
they reduce to the same conventional Fokker-Planck equation when one turns to the continuous-space limit. Rewriting eqs. \eqref{me} and \eqref{ME} for the lattices with an inter-site spacing $a$ and then setting $a \to 0$, one finds via a standard procedure  that $P_x(t)$  obeys
\begin{equation}
\label{FP}
\dot{P}_x(t) =  D \frac{d}{d x} \left(\frac{d }{d x} P_{x}(t) + \beta P_{x}(t) \frac{d U_x}{d x} \right) \,,
\end{equation}
where the diffusion coefficient $D = W_0 a^2$ and $D = a^2/2 \delta t$ for the Model I and the Model II, respectively. 
Both diffusion coefficients are assumed to be finite in the limit $a \to 0$, $W_0 \to \infty$ and $\delta t \to 0$.
In this limit, 
the expressions \eqref{dens1} and \eqref{dens2} for the probabilities of being at site $x$ at time $t \to \infty$ in both Models coincide and are given by the stationary solution of Eq. \eqref{FP} in a box of size $L$,
\begin{equation}
P_x(t = \infty) =\frac{ \exp\left(- \beta U_x\right)}{ \int^L_0 dx \exp\left(- \beta U_x\right)} \,.
\end{equation}

\begin{figure}
	\begin{center}
		\includegraphics[scale=0.33]{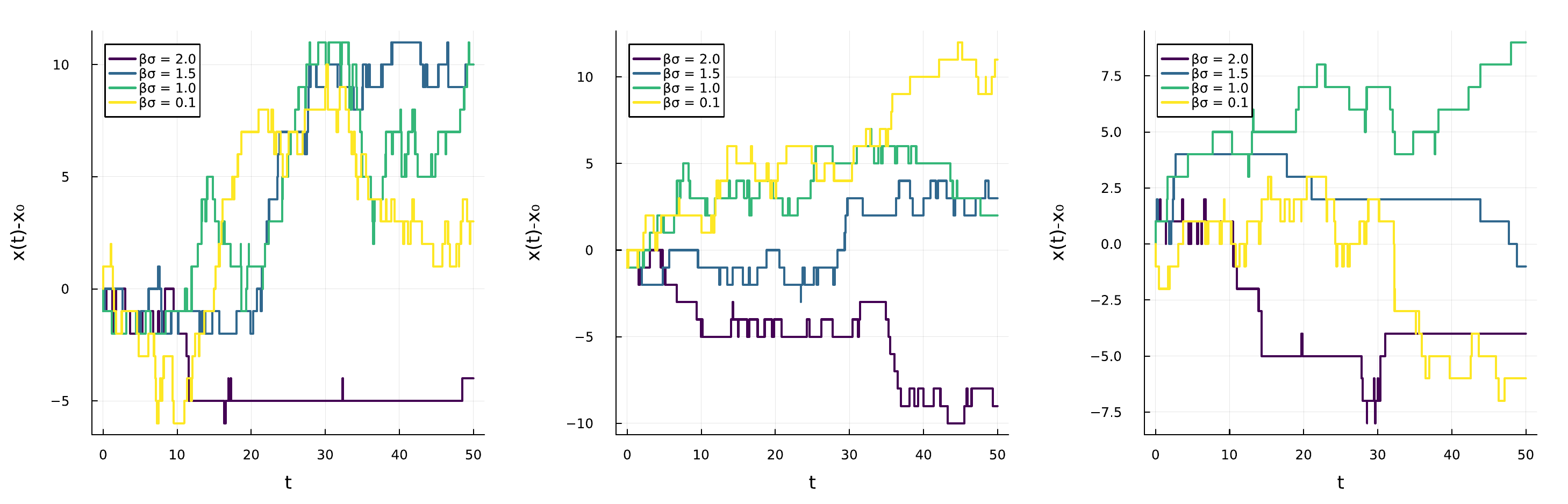}
	\end{center}
	\caption{Four individual short trajectories for the Model I (left panel), Model II (middle panel) and Model III (right panel) for four values of the parameter $\beta \sigma$ (see the inset).} 
	\label{fig:0}
\end{figure}

At the same time, given the difference between the master equations \eqref{me} and \eqref{ME}, 
we may expect that the behavior of the above-mentioned 
disorder-dependent random variables for finite inter-site spacing $a$ will be different for the two models under study. In Fig. \ref{fig:0}, in the left and middle panels, we depict four trajectories generated using the standard Gillespie algorithm \cite{gil} for the Models I and II, respectively, for four values of the parameter $\beta \sigma$,
which indeed show quite a different behavior for the same values of $\beta \sigma$. The difference becomes more pronounced the larger $\beta \sigma$ is.

\subsection{Scenario III: Gaussian random trap model.}

In the third scenario, we use the so-called random trap model (see \cite{jpb,leticia,mont,bertin,bertin2,ralf10,ralf11,ralf12,aki1,aki2,ralf13} and \cite{bou}), which offers a simplified yet physically plausible framework for studying slow dynamics, aging and anomalous diffusion in disordered media. This includes systems such as  glasses, spin glasses, disordered solids or other complex systems characterized by rugged energy landscapes featuring multiple local minima of random depths.  
In this context,  we assume that each site $x$ hosts a trap with a depth $-U_x$, $U_x$ being a quenched, negative definite random variable with the Gaussian probability density function in Eq. \eqref{Gaussm}. The values  of the depths of traps at different sites are uncorrelated. The particle undergoes activated hopping motion on this lattice and the time it spends in the trap at site $x$ is governed  by the Arrhenius law, meaning that this time is proportional to $\exp(-\beta U_x)$.

The time evolution of the probability $P_x = P_x(t)$ of finding the particle at site $x$ at a given time instant $t$ follows the master equation \eqref{me} with the transition rates
\begin{align}
\label{rates3}
W_{x,x \pm 1} = W_0 \, e^{\beta U_x} = W_0 \phi_x^2 \,.
\end{align}
Notably, the transition rates in this framework depend solely on the depth of the trap at the host site $x$ and are unaffected by the values $U_{x\pm1}$ that characterize the traps at the neighboring target sites. Such a model also satisfies the detailed balance and on a finite periodic chain with $N$ sites, $x = 1, 2, \ldots, N$,
the position probability $P_x$ approaches an equilibrium value as defined in Eq. \eqref{dens1}, mirroring the behavior observed in the Model I. In contrast, in the continuous-space limit $P_x$ does not obey the Fokker-Planck equation in Eq. \eqref{FP}. Instead, it follows a Fokker-Planck equation with a position-dependent diffusion coefficient. Four individual particle's trajectories generated for the discrete-space Gaussian random trap model are depicted in the right panel in Fig. \ref{fig:0} and show a markedly different behavior for sufficiently large values of $\beta \sigma$ as compared to the one observed in Models I and II.

\section{The realization-dependent currents and the resistances}
\label{sec:3}

Consider a finite chain of sites $x=0,1,2, \ldots, N$ with a fixed
sequence of $U_x$-s.  At site $x = 0$ we introduce a source of particles which 
maintains a constant occupation of this site, $P_0 = c_0$, while at the opposite extremity of the chain, i.e., at $x= N$,  
we place a perfect sink which removes particles from the system, $P_N = 0$. 
Such a system evolves towards a steady-state 
with a constant current $j_N$ to the sink, which is the following functional of a given sequence of $U_x$-s (see \cite{kalaj} and \ref{A}  for more details) :
\begin{align}
\begin{split}
\label{jN}
j_N^{(I)} = \frac{W_0 c_0}{\tau_N^{(I)}} \,, \quad
j_N^{(II)}  = \frac{c_0}{2 \delta t \tau_N^{(II)}} \,, \quad
j_N^{(III)}   =    \frac{W_0 c_0}{\tau_N^{(III)}}  \,,
\end{split}
\end{align}  
where the superscript indicates the dynamical scenario, while $\tau_N^{(I)}$, $\tau_N^{(II)}$ and $\tau_N^{(III)}$ are the corresponding to each scenario realization-dependent resistances of a finite chain. They are explicitly given by (see   \ref{A}) 
\begin{align}
\begin{split}
\label{taup}
\tau^{(I)}_N = \frac{H_{0,N}}{\phi_0^2} \,, \quad
\tau_N^{(II)} = \frac{H_{0,N}}{\phi_0 \phi_1} \,, \quad
\tau_N^{(III)} = \frac{N}{\phi_0^2} \,,
\end{split}
\end{align}
where $H_{0,N}$ is the following functional
\begin{equation}
\label{H}
H_{0,N} = \phi_0 \phi_1+ \phi_1 \phi_2 +  \phi_2 \phi_3 + \phi_3 \phi_4 + \ldots + \phi_{N-2} \phi_{N-1} + \phi_{N - 1} \phi_N  \,.
\end{equation}
In the latter expression the subscript $0$ corresponds to the first site in the sequence, i.e.,  $x=0$, 
while the subscript $N$ - to the last site $x = N$.

\subsection{The moments of resistances}

The first two moments of the realization-dependent resistances of the discrete-space finite chains can be straightforwardly calculated  in the leading in the large $N$ limit order, to give
 \begin{align}
\begin{split}
\label{momtau1}
&\Big \langle \tau_N^{(I)} \Big \rangle = \left \langle \phi \right \rangle^2 \left \langle \frac{1}{\phi^2} \right \rangle N + O(1) = \exp\left(3 \beta^2 \sigma^2/4\right) N + O(1) \,,\\
&\Big \langle \left(\tau_N^{(I)}\right)^2 \Big \rangle =  \left \langle \phi \right \rangle^4 \left \langle \frac{1}{\phi^4} \right \rangle^2 N^2 + O(N) 
= \exp\left(5 \beta^2 \sigma^2/2\right) N^2 + O(N) \,,
\end{split}
\end{align}
\begin{align}
\begin{split}
\label{momtau2}
&\Big \langle \tau_N^{(II)} \Big \rangle = \left \langle \phi \right \rangle^2 \left \langle \frac{1}{\phi} \right \rangle^2 N + O(1) = \exp\left(\beta^2 \sigma^2/2\right) N + O(1) \,,\\
&\Big \langle \left(\tau_N^{(II)}\right)^2 \Big \rangle =  \left \langle \phi \right \rangle^4 \left \langle \frac{1}{\phi^2} \right \rangle^2 N^2 + O(N) 
= \exp\left(3 \beta^2 \sigma^2/2\right) N^2 + O(N) \,,
\end{split}
\end{align}
and
\begin{align}
\begin{split}
\label{momtau3}
&\Big \langle \tau_N^{(III)} \Big \rangle = \left \langle \frac{1}{\phi^2} \right \rangle N =  \exp\left(\beta^2 \sigma^2/2\right) \left(1+ \erf\left(\beta \sigma/\sqrt{2}\right)\right) N  \,,\\
&\Big \langle \left(\tau_N^{(III)}\right)^2 \Big \rangle =   \left \langle \frac{1}{\phi^4} \right \rangle N^2  
= \exp\left( 2 \beta^2 \sigma^2 \right) \left(1 + \erf\left(\sqrt{2} \, \beta \sigma \right) \right)  N^2 \,,
\end{split}
\end{align}
where the symbols $O(1)$ and $O(N)$ signify that the omitted terms are independent of $N$ and are linear with $N$, respectively, while $\erf(x)$ denotes the error function. In Eqs. \eqref{momtau1} and \eqref{momtau2} the averaging is performed with respect to the probability density function \eqref{Pphi}, while the moments of $\tau_N^{(III)}$ are evaluated by using the analogous function in Eq. \eqref{Pphim}. Note that the moments of the resistances $\tau_N^{(I)}$ and $\tau_N^{(II)}$ exhibit a super-Arrhenius dependence on the temperature for any value of the parameter $\beta \sigma$, 
while the moments of $\tau_N^{(III)}$ show the super-Arrhenius behavior only for $\beta \sigma \to \infty$, in  which limit $\Big \langle \tau_N^{(III)} \Big \rangle \simeq 2 \exp(\beta^2 \sigma^2/2) N$.  
For small values of $\beta \sigma$, in the Taylor series
 expansion of $\tau_N^{(III)}$ the first subdominant term is linear with $\beta \sigma$, 
 \begin{align}
\Big \langle \tau_N^{(III)} \Big \rangle \simeq \left(1 + \sqrt{\frac{2}{\pi}} \, \beta \sigma + \frac{\beta^2 \sigma^2}{2} + O\left(\beta^3 \sigma^3\right) \right) N \,.
\end{align}

\begin{figure}
	\begin{center}
		\includegraphics[scale=0.348]{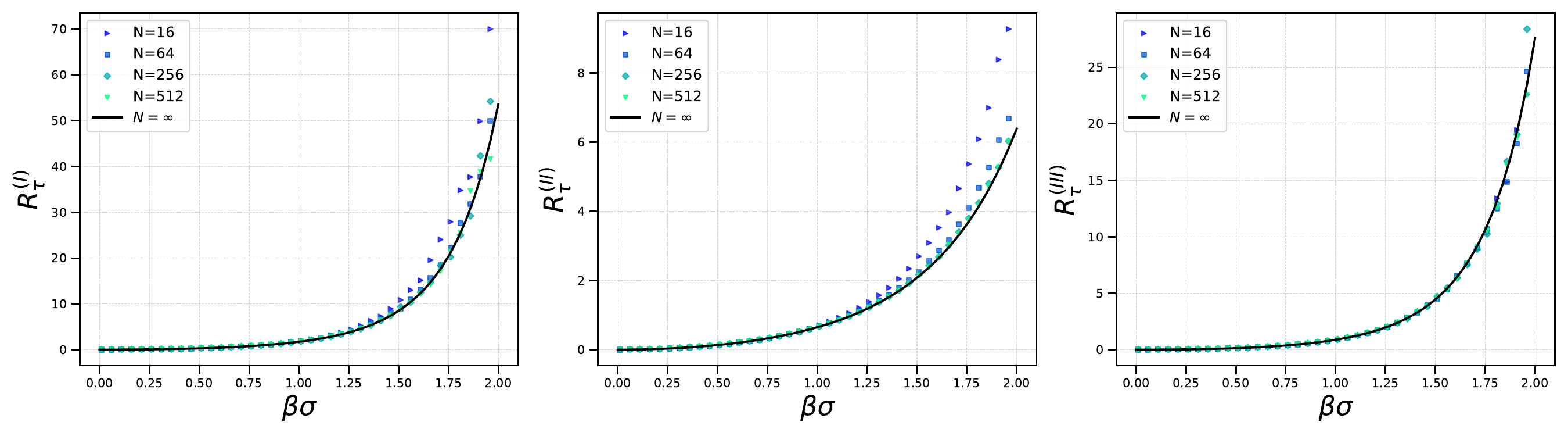}
	\end{center}
	\caption{The relative variances $R_{\tau}$ of the resistances $\tau_N$ in the Models I (left panel), II (middle panel)  and III (right panel) as functions of $\beta \sigma$. Solid curves are our analytical predictions in Eqs. \eqref{Rtau} to \eqref{Rtau3} for $N = \infty$, while the symbols depict $R_{\tau}$ evaluated by a numerical averaging of $\tau_N$ and $\tau^2_N$ for finite values of $N$ (see the insets).} 
	\label{fig:R}
\end{figure}

From Eqs. \eqref{momtau1} and \eqref{momtau2} we find that the corresponding 
relative variances in Eq. \eqref{R} obey
\begin{align}
\label{Rtau}
&R_{\tau^{(I)}} = \Big \langle \left(\tau^{(I)}_N\right)^2 \Big \rangle/\Big \langle \tau^{(I)}_N \Big \rangle^2 - 1 = \exp\left(\beta^2 \sigma^2\right) - 1 + O(1/N) \,,\\
\label{Rtau2}
&R_{\tau^{(II)}} = \Big \langle \left(\tau^{(II)}_N\right)^2 \Big \rangle/\Big \langle \tau^{(II)}_N \Big \rangle^2 - 1 = \exp\left(\beta^2 \sigma^2/2\right) - 1 + O(1/N) \,, 
\end{align}
while for the Model III the relative variance is independent of $N$ and is given by a more complicated expression of the form
\begin{align}
	\label{Rtau3}
R_{\tau^{(III)}} = \Big \langle \left(\tau^{(III)}_N\right)^2 \Big \rangle/\Big \langle \tau^{(III)}_N \Big \rangle^2 - 1 = \frac{\left(1 + \erf\left(\sqrt{2} \, \beta \sigma \right) \right)}{\left(1+ \erf\left(\beta \sigma/\sqrt{2}\right)\right)^2} \exp\left(\beta^2 \sigma^2\right) - 1 \,.
\end{align}
The above Eqs. \eqref{Rtau} and \eqref{Rtau2} imply that the resistances in  Models I and II are \textit{not self-averaging} in the limit $N \to \infty$. The resistance in the Model III is not self-averaging for any value of $N$.
Moreover, in all three Models the relative variances $R_{\tau}$ show an unbounded growth with $\beta \sigma$, which signifies 
that fluctuations become more important the larger the disorder is. The behavior of $R_{\tau}$ for all three Models is depicted in Fig. \ref{fig:R}.
Below we will show that such a lack of self-averaging is essentially the boundary effect.

\subsection{The limit distributions of the resistances and their typical values.}

To get an idea why the resistances are not self-averaging in all three Models, we 
seek their probability density functions. We start with Model I.  The limiting for  $N \to \infty$ form of the probability density function of $\tau^{(I)}_N$ can be evaluated following the standard approach: We first 
consider a reduced random variable
\begin{equation}
\label{z}
\overline{\tau}_{(I)} = \frac{\tau^{(I)}_N}{N} \,,
\end{equation}
and focus on the form of its moment-generating function in the limit $N \to \infty$,
\begin{equation}
\label{mg1}
\Phi(\overline{\tau}_{(I)}) = \lim_{N \to \infty} \Big \langle \exp\left(- \lambda \overline{\tau}_{(I)}\right) \Big \rangle \,, \quad \lambda \geq 0 \,,
\end{equation} 
where the angle brackets denote the averaging over all random variables $\phi_x$.
Once $\Phi(\overline{\tau}_{(I)})$ is known, the probability density function $P(\overline{\tau}_{(I)})$ obtains by a mere inversion of the Laplace transform.

We  formally split the averaging over all sites in 
Eq. \eqref{mg1} into the averaging over the variables $\phi_x$ for all the "bulk" sites $ x= 1,2,3, \ldots, N$,   
and the averaging over the variable $\phi_0$, such that  
\begin{align}
\begin{split}
\label{zut}
\Phi(\overline{\tau}_{(I)}) &= \lim_{N \to \infty} \Big \langle \Big \langle \exp\left( - \lambda \overline{\tau}_{(I)}\right) \Big \rangle_{\rm bulk} \Big \rangle_{0}\\
&= \lim_{N \to \infty} \left \langle \left \langle \exp\left( - \frac{\lambda}{N \phi_0^2} H_{0,N} \right) \right \rangle_{\rm bulk} \right \rangle_{0}  \,, 
\end{split}
\end{align}
We notice next that $H_{0,N}$ in Eq. \eqref{H} is self-averaging. Indeed, we have
\begin{align}
\begin{split}
&\Big \langle  H_{0,N}  \Big \rangle_{\rm bulk} = \langle \phi \rangle^2 N + O(1) \,, \\
&\Big \langle  H_{0,N}^2  \Big \rangle_{\rm bulk} = \langle \phi \rangle^4 N^2 + O(N) \,,
\end{split}
\end{align} 
such that 
\begin{equation}
R_H = O(1/N) 
\end{equation}
and hence, it vanishes in the limit $N \to \infty$. This implies that the Jensen gap also vanishes in this limit and
\begin{align}
\Phi(\overline{\tau}_{(I)}) &=  \left \langle \exp\left( - \frac{\lambda \langle \phi \rangle^2}{\phi_0^2}\right)  \right \rangle_{0} \,.
\end{align}
Hence, the probability density function of $\overline{\tau}_{(I)}$ obeys
\begin{align}
\begin{split}
\label{aa}
P(\overline{\tau}_{(I)}) &=  \left \langle \delta\left(\overline{\tau}_{(I)} - \frac{e^{\beta^2 \sigma^2/4}}{\phi_0^2}\right)  \right \rangle_{0} \\
&= \frac{1}{\sqrt{2 \pi \beta^2 \sigma^2} \left(\overline{\tau}_{(I)}\right)^{3/4}} \exp\left( - \frac{\beta^2 \sigma^2}{32} - \frac{\ln^2\left(\overline{\tau}_{(I)}\right)}{2 \beta^2 \sigma^2}\right) \,, \quad 0 \leq \overline{\tau}_{(I)} < \infty \,.
\end{split}
\end{align}
One can straightforwardly check that the first two moments of this function coincide with those defined in Eq. \eqref{momtau1}.

The corresponding probability density function of the resistance in the Model II can be calculated using essentially the same kind of arguments. In doing so, we find
\begin{align}
\begin{split}
\label{bb}
P(\overline{\tau}_{(II)}) &= \left \langle \delta\left(\overline{\tau}_{(II)} - \frac{e^{\beta^2 \sigma^2/4}}{\phi_0 \phi_1}\right)  \right \rangle_{x=0,1} \\
&= \frac{1}{\sqrt{\pi \beta^2 \sigma^2} \left(\overline{\tau}_{(II)}\right)^{1/2}} \exp\left(- \frac{\beta^2 \sigma^2}{16} - \frac{\ln^2 \left(\overline{\tau}_{(II)}\right)}{\beta^2 \sigma^2} \right) \,,  \quad 0 \leq \overline{\tau}_{(II)} < \infty \,.
\end{split}
\end{align}
It is easy to find that the first two moments of the latter probability density function are given precisely by Eq. \eqref{momtau2}. Note, as well, that the probability density functions in Eqs. \eqref{aa} and \eqref{bb} have the form of the log-normal distribution as the parental form in Eq. \eqref{Pphi}, and differ from it (and between themselves) only in the pre-exponential algebraic dependence on the reduced resistance and by the normalizations. 

Lastly, for the Model III the probability density function of the variable $\overline{\tau}_{(III)} = \tau^{(III)}_N/N$ is calculated very straightforwardly. Using Eq. \eqref{Pphim}, we find the log-normal distribution of the form
\begin{align}
\begin{split}
\label{cc}
P(\overline{\tau}_{(III)}) &= \sqrt{\frac{8}{\pi \beta^2 \sigma^2}} \int^1_0 \frac{d\phi}{\phi} \, \delta\left(\overline{\tau}_{(III)} - \frac{1}{\phi^2}\right) \exp\left(- \frac{2}{\beta^2 \sigma^2} \ln^2(\phi)\right) \\
& = \sqrt{\frac{2}{\pi \beta^2 \sigma^2}} \frac{1}{\overline{\tau}_{(III)}} \exp\left(- \frac{\ln^2\left(\overline{\tau}_{(III)}\right)}{2 \beta^2 \sigma^2}\right) \,,\quad 1\leq \overline{\tau}_{(III)} < \infty \,.
\end{split}
\end{align}

\begin{figure}
	\begin{center}
		\includegraphics[scale=0.348]{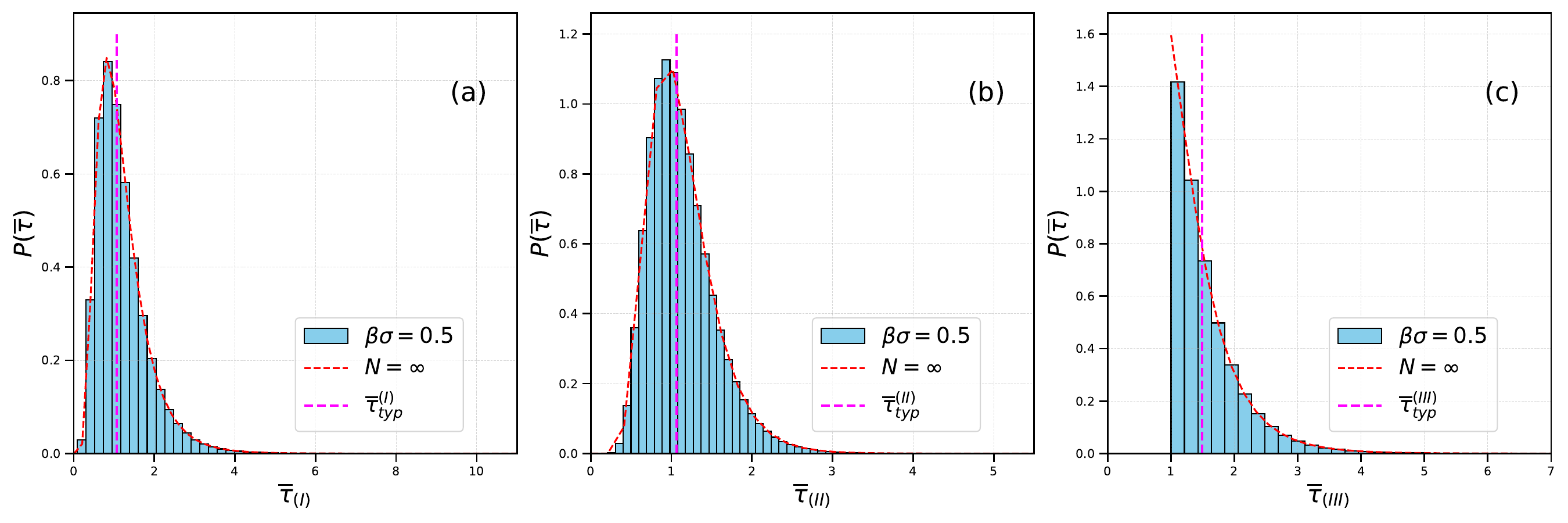}
	\end{center}
	\caption{The probability density functions of the reduced resistance $\overline{\tau}$ for the Models I (panel (a)), II (panel (b)) and III (panel (c)). The histrograms present the results of a numerical analysis of the random variable $\overline{\tau}$. The enveloping dashed (red) curve depicts our analytical predictions in  Eqs. \eqref{aa}, \eqref{bb} and \eqref{cc}, respectively. Vertical dashed lines indicate the typical values of $\overline{\tau}$ (see Eqs. \eqref{tautyp} and \eqref{tautyp3}). } 
	\label{fig:Ptau}
\end{figure}

Figure \ref{fig:Ptau} illustrates the probability density functions given in Eqs. \eqref{aa}, \eqref{bb}, and \eqref{cc}, along with numerically constructed histograms for the random variable 
$\overline{\tau}$. The comparison shows an excellent agreement between our analytical predictions and the numerical results.

We close this subsection with the following three remarks:\\ -- The resistances in  the Models I and II are not self-averaging in the limit $N \to \infty$, while the resistance in the Model III is not self-averaging for any $N$. The resistances exhibit significant fluctuations with the relative variances becoming very large when $\beta \sigma$ is large.\\ -- The lack of self-averaging is solely due to fluctuations of the on-site potentials at the site $x=0$ (for Models  I and III) and at the sites $x=0,1$  (for Model II). The contributions of the sites in the bulk of the chain are self-averaging for Models I and II, and is constant, i.e., non-fluctuating for the Model III.\\
-- Given the lack of self-averaging of the resistances, it is instructive to consider their typical behavior which should be observed for most of realizations of sequences of $U_x$ and can be accessed by calculating the averaged logarithm of $\overline{\tau}$. Using the expressions \eqref{aa} and \eqref{bb}, we 
determine this property to get the following super-Arrhenius function of the temperature for the Models I and II:
\begin{align}
\label{tautyp}
\overline{\tau}^{\rm (I \, and \, II)}_{\rm typ} = \exp\left(\Big \langle \ln(\overline{\tau}^{\rm (I \, and \, II)}) \Big \rangle\right) = \exp\left(\frac{\beta^2 \sigma^2}{4}\right)  \,.
\end{align}
Unexpectedly, the typical value of $\overline{\tau}$ is the \textit{same} for both Models, despite the fact that their moments grow with $\beta \sigma$ at different rates.  The typical value is smaller than the averaged ones, which signifies that the averaged values are supported by some atypical realizations of sequences of the on-site potentials.  

Concurrently, for the Model III we find
\begin{align}
\label{tautyp3}
\overline{\tau}^{(III)}_{\rm typ} = \exp\left(\Big \langle \ln(\overline{\tau}^{(III)}) \Big \rangle\right) = \exp\left(\sqrt{\frac{2}{\pi}} \, \beta \sigma\right)  \,.
\end{align}
Surprisingly, the typical resistance of a finite chain in the Gaussian random trap model exhibits the Arrhenius dependence on the temperature, in striking contrast to the behavior observed in Models I and II.
The typical values of $\overline{\tau}$ are indicated in Fig. \ref{fig:Ptau} by vertical dashed lines. We observe that for the Models I and II the values of $\overline{\tau}_{\rm typ}$ appear very close to the most probable values of this variable, while for the Model III it is also close to the left edge of the support  at which the probability density function attains its maximal value.

\subsection{Probability currents}

Consider Eqs. \eqref{jN} and conveniently rewrite them in terms of the reduced random variables $\overline{\tau}$:
\begin{align}
\begin{split}
\label{jNr}
j_N^{(I)}  = \left(\frac{W_0 c_0}{N}\right)  \frac{1}{\overline{\tau}_{(I)}  }       \,, \quad 
j_N^{(II)}  = \left(\frac{c_0}{2 \delta t N}\right)  \frac{1}{\overline{\tau}_{(II)}  }  \,,\quad
j_N^{(III)}  = \left(\frac{W_0 c_0}{N}\right)  \frac{1}{\overline{\tau}_{(III)}  }    \,,
\end{split}
\end{align}  
where the terms in the brackets can be identified as currents in the absence of disorder. 
We find next from the expressions \eqref{aa}, \eqref{bb} and \eqref{cc} that 
the large-$N$ asymptotic behavior of  the moments of the currents in Models I and II  is given explicitly by
\begin{align}
\begin{split}
\label{jNrm}
\Big \langle \left(j_N^{(I)}\right)^q \Big \rangle & = \left(\frac{W_0 c_0}{N}\right)^q  e^{q (2 q - 1) \beta^2 \sigma^2/4}       \,, \\
\Big \langle \left(j_N^{(II)}\right)^q \Big \rangle & = \left(\frac{c_0}{2 \delta t N}\right)^q  e^{q (q - 1) \beta^2 \sigma^2/4}  \,, 
\end{split}
\end{align} 
while for the Model III the exact result, valid for all $N$, follows
\begin{align}
\begin{split}
\label{jNrm3}
\Big \langle \left(j_N^{(III)}\right)^q \Big \rangle & = \left(\frac{W_0 c_0}{N}\right)^q  e^{q^2 \beta^2 \sigma^2/2}   \erfc(q \beta \sigma/\sqrt{2}) \,,
\end{split}
\end{align} 
where $\erfc(x)$ is the complementary error function. From the above expressions we readily find that  
the relative variances of the currents obey
\begin{align}
\begin{split}
\label{Rj}
R_{j^{(I)}} & = \exp\left(\beta^2 \sigma^2\right)  - 1       \,, \\
R_{j^{(II)}} & =  \exp\left(\beta^2 \sigma^2/2\right) - 1  \,,\\
R_{j^{(III)}} & =  \frac{\erfc(\sqrt{2} \, \beta \sigma)}{\erfc^2(\beta \sigma/\sqrt{2})}  \exp\left(\beta^2 \sigma^2\right) - 1 \,,
\end{split}
\end{align}  
which implies that the currents, likewise the resistances, are not self-averaging in all three Models. Moreover, the corresponding relative variances of the current are monotonically increasing functions of the parameter $\beta \sigma$.

\begin{figure}
	\begin{center}
		\includegraphics[scale=0.34]{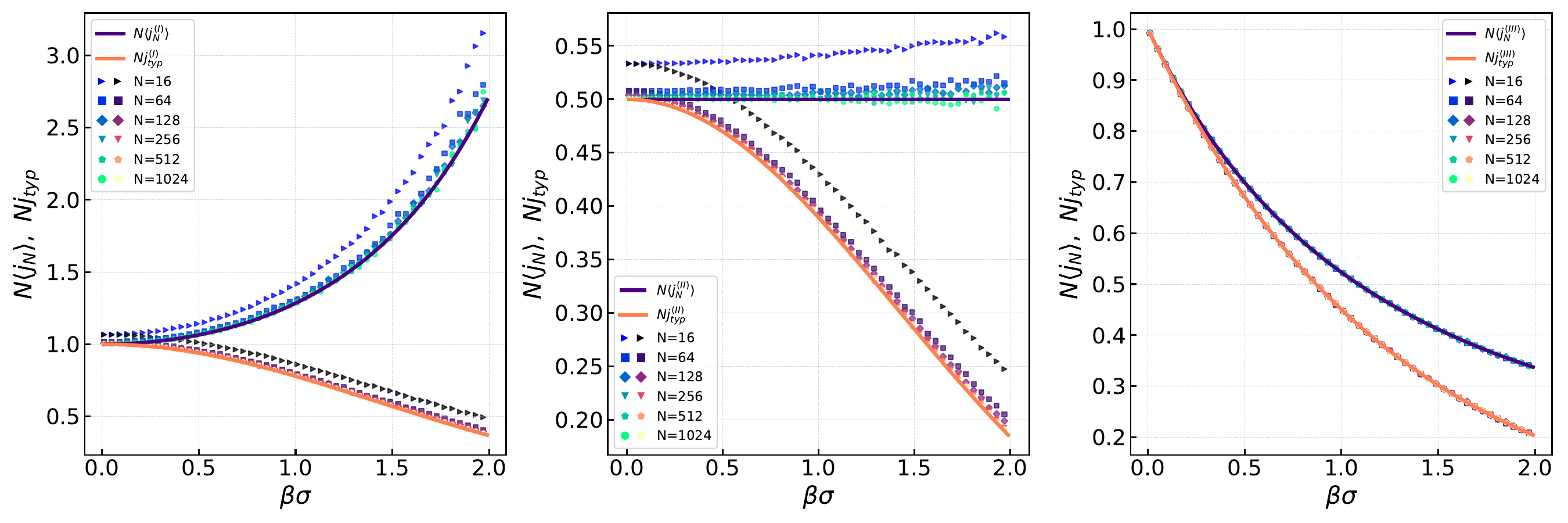}
	\end{center}
	\caption{The disorder-averaged versus typical currents (multiplied by $N$) through a finite interval with $N$ sites for the Models I (left panel), II (middle panel) and III (right panel). The black  solid curves depict our analytical predictions for the averaged currents in Eqs. \eqref{jNrm} and \eqref{jNrm3} (with $q = 1$). The red solid curves depict the typical currents in Eqs.  \eqref{typj} and \eqref{typj3}. Symbols present the results of a numerical averaging of $j_N$ in Eqs. \eqref{jNr} and the results for an exponentiated averaged $\ln j_N$ (see the inset). } 
	\label{fig:jN}
\end{figure}

Upon a closer look on the expressions \eqref{jNrm} and \eqref{jNrm3} we notice that the disorder-average currents 
in the three Models under study exhibit distinctly different behavior as functions of $\beta \sigma$ (see  Fig. \ref{fig:jN}):  in the Model I
the disorder-averaged current $\Big \langle j_N^{(I)} \Big \rangle$ is a monotonically \textit{increasing} function of  $\beta \sigma$. This is apparently not very counter-intuitive at the first glance. Indeed, one can imagine such realizations of disorder  in  which, upon an increase  
 of $\beta \sigma$, we increase the magnitude of  random forces which push the particle to move along the chain towards the site $x = N$. Concurrently, the disorder-averaged current  
in the Model II  is \textit{independent} of disorder and has precisely the form of the standard Fickian result for the current  in a homogeneous system. This means that evidently in the Model II some very particular, anomalous realizations of disorder control the value of the disorder-averaged current.   A  similar behavior has been observed previously in \cite{kalaj} for a somewhat different model. 
On the contrary, in Model III the current is a monotonically  \textit{decreasing}  function of $\beta \sigma$.  In the leading order in the limit $\beta \sigma \to \infty$ the disorder-averaged current in the Gaussian random trap model exhibits a power-law dependence on disorder of the form
 \begin{align}
 \label{currentM3}
\Big \langle j_N^{(III)} \Big \rangle & \simeq \left(\frac{W_0 c_0}{N}\right)  \sqrt{\frac{2}{\pi}} \frac{1}{\beta \sigma} \,.
\end{align}
In fact, this is again not a counter-intuitive behavior. Indeed, increasing the value of $\beta \sigma$ we increase the depths of traps such that the retention times become larger.

It is important also to compare the behavior of averaged properties against the typical behavior, which should be observed for a majority  of realizations of disorder. The typical behavior of the currents can be accessed by merely noticing that, up to 
the prefactors, in each Model $\ln j_N = - \ln \tau_N$, and by using then our Eqs. \eqref{tautyp} and \eqref{tautyp3}. 
We find that for the Models I and II the typical behavior follows universally
\begin{equation}
\label{typj}
j_{typ}^{\rm (I \, and \, II)} = \exp\left(\Big \langle \ln j^{\rm (I \, and \, II)}_N \Big \rangle\right) \sim e^{- \beta^2 \sigma^2/4} \,,
\end{equation}
i.e., the typical current in these two Models is characterized by the same \textit{decreasing}, super-Arrhenius  
function of disorder and the temperature and therefore, is
much smaller than the averaged currents in eqs. \eqref{jNrm}. Such a strong "reduction" of the current is apparently due to the fact that in typical realizations of disorder a particle will always get localized between two sites at which sufficiently large forces point in the opposite directions.    
 Further on, we find that in  the Model III the typical current obeys
\begin{align}
\label{typj3}
j_{typ}^{(III)} = \exp\left(\Big \langle \ln j^{(III)}_N \Big \rangle\right) \sim  \exp\left(- \sqrt{\frac{2}{\pi}} \, \beta \sigma\right) \,,  
\end{align}
i.e.,  is an exponential Arrhenius function of $\beta \sigma$, in place of the power-law dependence predicted in Eq. \eqref{currentM3}. The typical current in this Model also appears to be much smaller than the averaged one.  
Expressions \eqref{typj} and \eqref{typj3}  highlight a significant difference between the typical and averaged behavior showing that actually in all scenarios the averaged behavior is supported by atypical realizations of disorder. Behavior of the typical currents is confronted against the one of the averaged currents in Fig. \ref{fig:jN}, which substantiates the above discussion.

\subsection{Sample-to-sample fluctuations of the resistances and currents}

On example of Model I we discuss the behavior of yet another characteristic property which highlights the role of sample-to-sample fluctuations 
showing how \textit{disproportionally} different, or conversely, similar 
can be the values of resistances (or currents) observed in two different samples. 
Following Refs. \cite{7,8}, 
we consider two finite chains of the same length $N$ with two different realizations of the sequences of the on-site potentials $U_x$. 
We assume next that  
the values of the reduced resistances defined in Eq. \eqref{z} are $\overline{\tau}_1$ and $\overline{\tau}_2$, respectively, and the corresponding values of the currents are $j_1$ and $j_2$. 
For brevity, we skip the subscript $(I)$ in what follows as well as the subscript $N$.  
Then, we construct a random variable 
\begin{align}
\label{omega}
\omega = \frac{\overline{\tau}_1}{\overline{\tau}_1 + \overline{\tau}_2} = \frac{j_2}{j_1 + j_2} \,,
\end{align}
which defines a relative contribution of either of the resistances (currents) to their sum. We seek next the 
probability density function $P(\omega)$ of this random variable with the support on $(0,1)$.  Note that because the probability density functions of the resistances have very similar forms (see Eqs. \eqref{aa}, \eqref{bb} and \eqref{cc}), we do not expect any significant difference of the behavior of $P(\omega)$ for the three models under study and therefore, constrain our analysis for the Model I only.

Clearly, the first moment of this random variable $\langle \omega \rangle = 1/2$, by symmetry, but this result can be meaningful or misleading depending on the shape  
 of  $P(\omega)$. If $P(\omega)$ is unimodal and centered at $\omega = 1/2$, this will signify that
it is most likely that $\overline{\tau}_1 \approx \overline{\tau}_2$ and sample-to-sample fluctuations are not very important. If, on the contrary, $P(\omega)$ is bimodal and has a minimum at $\omega = 1/2$, this will imply that most likely $\overline{\tau}_1$ and  $\overline{\tau}_2$ are disproportionally different, despite the fact that averaged value of $\omega$ is equal to $1/2$. 

\begin{figure}[ht]
\centering
\includegraphics[width=70mm,angle=0]{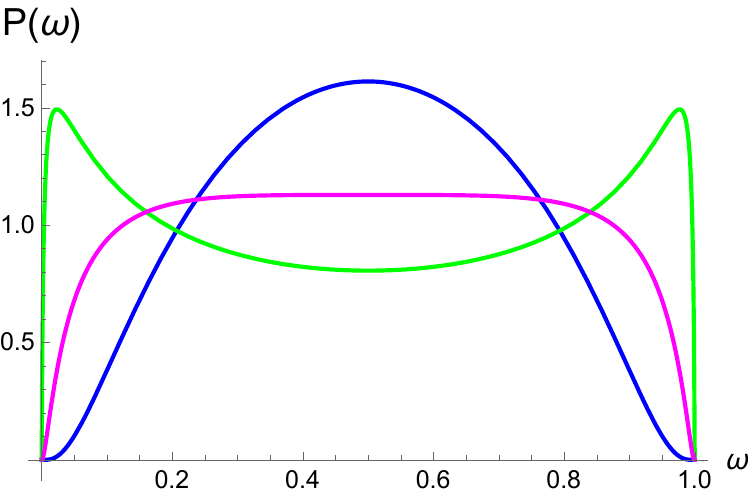}  
\caption{Sample-to-sample fluctuations. Probability density function  $P(\omega)$ in Eq. \eqref{Pomega} as function of $\omega$ for $\beta \sigma = 0.1$ (blue curve),  
 $\beta \sigma = 1$ (magenta curve) and $\beta \sigma = 1.2$ (green curve). }
\label{fig:1} 
\end{figure}

In order to calculate  $P(\omega)$ we consider first its moment-generating function
\begin{align}
\begin{split}
\Phi_{\omega} &= \Big \langle \exp\left( - \lambda \omega \right) \Big \rangle  \\
&=\int^{\infty}_0 d\overline{\tau}_1 \int^{\infty}_0 d\overline{\tau}_2 \, \exp\left(- \lambda \frac{\overline{\tau}_1} {\overline{\tau}_1 + \overline{\tau}_2}\right)  \, P(\overline{\tau}_1)  \, P(\overline{\tau}_2) \,,  
\end{split}
\end{align}
where $P(\overline{\tau})$ is defined in Eq. \eqref{aa}. Changing formally the integration variable $\overline{\tau}_1 \to \omega$, Eq. \eqref{omega}, we have
\begin{align}
\begin{split}
\Phi_{\omega} &= \int^{1}_0 \frac{d \omega}{(1 - \omega)^2} \, e^{- \lambda \omega} \int^{\infty}_0 \overline{\tau}_2  \, d\overline{\tau}_2   \,   P(\overline{\tau}_2) \, P\left(\frac{\omega}{1 - \omega} \, \overline{\tau}_2\right) \,,
\end{split}
\end{align}
from which we can readily read of the desired probability density function
\begin{align}
\begin{split}
\label{Pomega}
P(\omega) &=  \frac{1}{(1 - \omega)^2} \int^{\infty}_0 \overline{\tau}_2 \, d\overline{\tau}_2   \,  P(\overline{\tau}_2) \, P\left(\frac{\omega}{1 - \omega} \, \overline{\tau}_2\right) \\
& = \frac{1}{2 \sqrt{\pi \beta^2 \sigma^2} \omega (1 - \omega)} \, \exp\left( - \frac{\ln^2\left(\omega/(1 - \omega)\right)}{4 \beta^2 \sigma^2}\right) \,.
\end{split}
\end{align}
This probability density function is depicted in Fig. \ref{fig:1} for three values of the parameter $\beta \sigma$. We observe that for a relatively small disorder, i.e., for $\beta \sigma = 0.1$, $P(\omega)$ is unimodal and centered at $\omega = 1/2$. This signifies that for a small disorder it is indeed most likely that the values of resistances (or currents) observed in two randomly chosen samples are very close to each other. Upon an increase of $\beta \sigma$, for $\beta \sigma = 1$,  we encounter a different behavior when the central part of $P(\omega)$ becomes flat and $\omega = 1/2$ is no longer the most probable value. In this situation, (almost) any relation  between the two resistance becomes equally probable. Increasing $\beta \sigma$ further, we have that already for $\beta \sigma = 1.2$ the probability density function becomes bimodal with two maxima close to $\omega = 0 $ and $\omega = 1$. Here, $\omega = 1/2$ corresponds not to the most probable but to the \textit{least} probable value. This signifies that the resistance (or currents) observed in two sample are most likely disproportionally different and sample-to-sample fluctuations are very significant. 

\section{The realization-dependent splitting probability}
\label{sec:4}
Consider a one-dimensional chain composed of \( N + 1 \) sites labeled 
\( x = 0, 1, \ldots, N \). 
Suppose that the particle starts at site \( x = x_0 \) and performs a continuous-time random walk 
with site-dependent transition rates 
\( W_{x,\,x+1} \) (to the right, \( x \to x+1 \)) 
and \( W_{x,\,x-1} \) (to the left, \( x \to x-1 \)).
The sites \( x = 0 \) and \( x = N \) at the boundaries are absorbing, 
meaning that once the particle reaches either boundary, it remains there forever.
We  focus here on the so-called \emph{splitting probability} $E_-(x_0)$; 
that is, the probability that a particle starting from the site \( x_0 \)
is eventually absorbed at the left boundary \( x = 0 \) 
before ever reaching the right boundary \( x = N \).

In \ref{B} we show that the realization-dependent splitting probability admits the following compact representation,
\begin{equation}
\label{eq:sol_bella}
E_-(x_0) =
\frac{\displaystyle\tau_{N-x_0}^+}{\displaystyle\tau_{x_0}^- + \tau_{N-x_0}^+},
\end{equation}
where $\tau_{x_0}^+$ and $\tau_{N-{x_0}}^-$ are the  resistances of the intervals $(0,x_0-1)$  and $(x_0+1,N)$ from the left and from the right of the starting point $x_0$, respectively. These realization-dependent resistances are defined explicitly by rather lengthy expressions (see Eqs.  \eqref{eq:res1res2}) which we present in \ref{B}. An analogous expression for the continuous-space system can be found in \cite{d11}.

As shown in \ref{B}, the realization-dependent splitting probability for  Model III is trivially $E_-(x_0) = 1 - x_0/N$, i.e., it is not a fluctuating property and is the same as in homogeneous systems without disorder \cite{redner}.  Concurrently, for the Models I and II the splitting probabilities do depend on disorder and for both Models are given by
\begin{align}
    \label{eq:splitI&II}
    E_-(x_0)=\frac{H_{x_0,\,N}}{H_{0,N}}\,,
\end{align}
where the function $H_{x_0,\,N}$ is defined by Eq. \eqref{H} with the first index replaced by $x_0$.
A straightforward  analysis shows that in the joint limit  $N,\,x_0\to \infty$ with the ratio $\alpha=x_0/N$ kept fixed,
\[
\frac{1}{N}\sum_{x=0}^{N-1}\phi_x\phi_{x+1}
\;\xrightarrow[N\to\infty]{}\;
\langle \phi \rangle^2,
\qquad
\frac{1}{N-x_0}\sum_{x=x_0}^{N-1}\phi_x\phi_{x+1}
\;\xrightarrow[N,  x_0 \to\infty]{}\;
\langle \phi \rangle^2,
\]
almost surely. As a consequence, the realization-dependent splitting probability self-averages in this limit and converges to
\begin{equation}
\label{eq:asymp_split}
E_-(x_0) \simeq 1-\frac{x_0}{N},
\end{equation}
with vanishing sample-to-sample fluctuations. 

\begin{figure}
    \centering
    \includegraphics[width=0.9\linewidth]{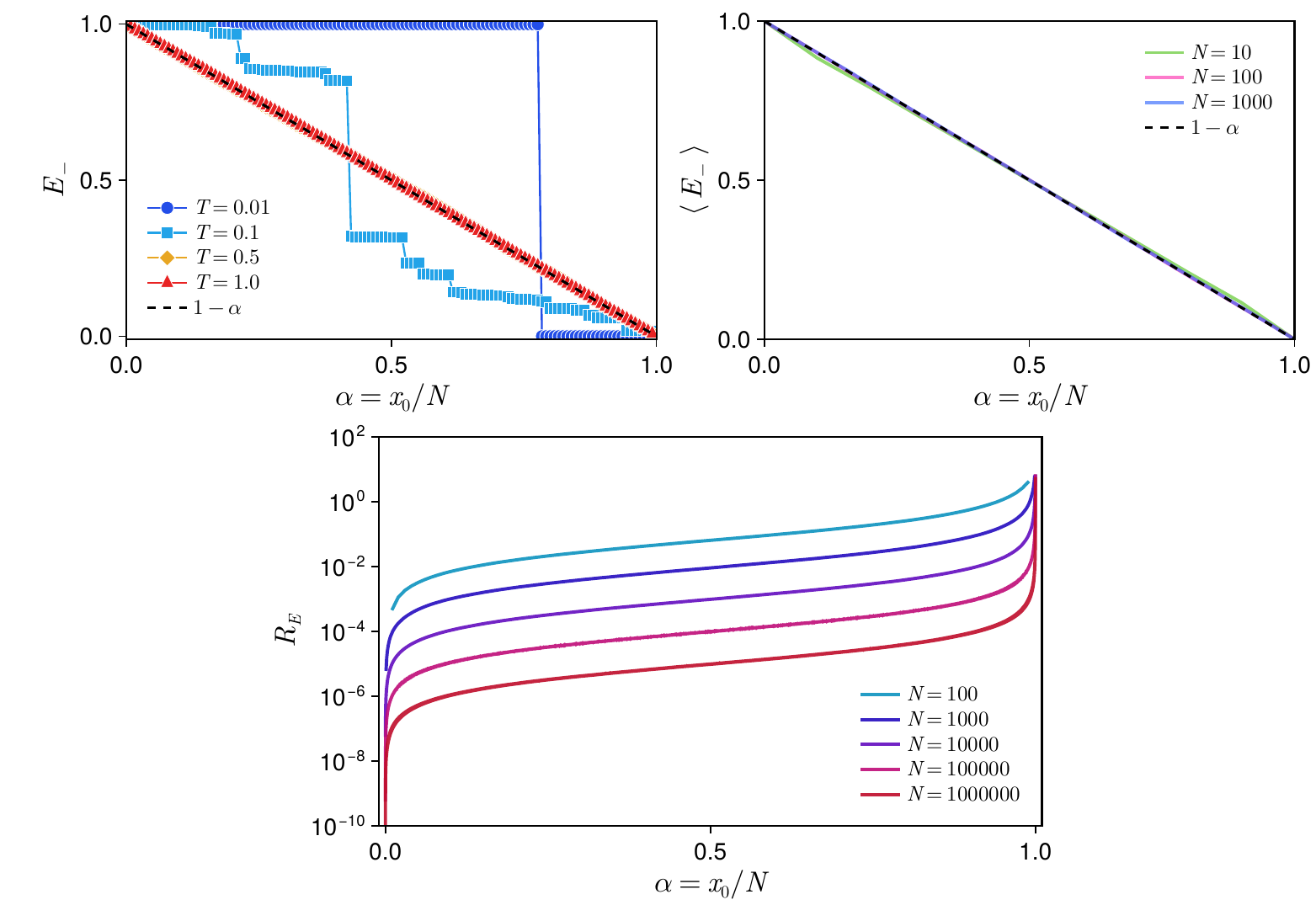}
    \caption{
The splitting probability $E_-(x_0)$ as function of $\alpha = x_0/N$.   Top row: The splitting probability in Eq. \eqref{eq:splitI&II} for a given realization of disorder on a chain with $N=10^5$ sites for different temperatures (left panel). 
Numerically-evaluated first moment of the splitting probability for several values of $N$ at fixed inverse temperature $\beta=10$ (right panel). In both panels the dashed line depicts the analytical prediction in Eq. \eqref{eq:asymp_split}. 
Bottom row: Relative variance $R_{E} = \mathrm{Var}(E_-(x_0))/\langle E_-(x_0)\rangle^2$  for $\beta=2$ and several values of $N$. 
}
    \label{fig5}
\end{figure}
In Fig. \ref{fig5} we show the splitting probability as a function of the scaled position $\alpha=x_0/N$. The top row (left panel) displays the realization-dependent $E_{-}(x_0)$ given by Eq. \eqref{eq:sol_bella}, evaluated for a single randomly generated disorder realization on a chain of $10^5$ sites, at several temperatures.
At the lowest temperature ($T=0.01$), the sum $H_{0,N}$ is entirely dominated by its largest term. As a result, the splitting probability takes the form of a step function: $E_{-}(x_0)=1$ up to a critical realization-dependent value of $\alpha$, beyond which it drops abruptly to zero.
When the temperature is increased by an order of magnitude, subdominant contributions to $H_{0,N}$ become relevant. Consequently, $E_{-}(x_0)$ develops a descending staircase-like structure as a function of $\alpha$, with each plateau corresponding to dominance of of different terms in the sum.
Upon further increase of the temperature, the realization-dependent fluctuations are progressively smoothed out. In this regime, the realization-dependent splitting probability becomes nearly indistinguishable from that of a homogeneous system, recovering the linear behavior $E_{-}(x_0)=1-\alpha$ \cite{redner}. Next, the top-right panel of Fig. \ref{fig5} shows the numerically evaluated disorder-averaged splitting probability for several values of $N$ at a fixed temperature. We find that all curves are nearly indistinguishable from the asymptotic prediction in Eq. \eqref{eq:asymp_split}, indicating that size-dependent corrections remain very small, even at the relatively low temperature $T=0.1$.
Lastly, the bottom row of Fig. \ref{fig5} displays the relative variance of the realization-dependent splitting probability as a function of $\alpha = x_0/N$ for several values of $N$ (see inset), at a fixed inverse temperature $\beta = 2$. Away from the boundaries $\alpha = 0$ and $\alpha = 1$, where $E_-$ equals $1$ and $0$, respectively, and hence $R_E \equiv 0$, the relative variance is nonzero for finite $N$ but decreases systematically as $N$ increases. In the limit $N \to \infty$, $R_E$ vanishes, indicating that the splitting probability becomes self-averaging.

\section{The realization-dependent mean first-passage times}
\label{sec:5}

Consider a finite chain with $N+1$ sites $x$, $x  = 0, 1, \ldots, N$, in which the site $x = N$ is the "target". A particle starts 
from the site at the left extremity of the chain, i.e.,  from $x = 0$, performs a random walk in a continuous-time between the nearest-neighboring sites of the chain and reaches for the first time the right extremity at some random time instant, which is called the first-passage time to $N$. The \text{mean} first-passage time $T_N$, with the averaging being performed over all possible realizations of the random walk process for a fixed realization of the transition rates $W_{x, x \pm 1}$, is an important 
characteristic property of random transport processes which has been amply analyzed in the past (see, e.g., \cite{kehr,bou,hughes} and \cite{katja2,redner,metz,olivier,denis}).  Here we follow an early work  \cite{weiss} in which an explicit expression for $T_N$ has been derived for the continuous-time and discrete-space master equation for the general form of the transition rates. 
Adapting this result to our notations, one has
\begin{align}
\begin{split}
\label{TN}
T_N &= \frac{\eta_N}{\theta_N} \sum_{j=0}^{N-1} \theta_{j} - \sum_{j=0}^{N - 1} \eta_j \,,  
\end{split}
\end{align}
where
\begin{align}
\begin{split}
\label{theta}
\theta_0 = 1 \,, \quad \theta_j &= 
\frac{W_{0,1} W_{1,2} W_{2,3} \ldots W_{j-1,j}}{W_{1,0} W_{2,1} W_{3,2} \ldots W_{j,j-1}} \,, \\
\eta_0 = 0 \,, \quad \eta_j &= \frac{1}{W_{j,j-1}} \Big[1 + \frac{W_{j-1,j}}{W_{j-1,j-2}} + \frac{W_{j-1,j} W_{j-2,j-1}}{W_{j-1,j-2} W_{j-2,j-3}}  \\
&+ \ldots + \frac{W_{j-1,j} W_{j-2,j-1} W_{j-3,j-2} \ldots W_{1,2}}{W_{j-1,j-2} W_{j-2,j-3} W_{j-3,j-4} \ldots W_{1,0}} \Big] \,.
\end{split}
\end{align}
For the Model I the rates are defined in Eq. \eqref{rates}, while for the Model II we have $W_{x,y} = p_{x,y}/\delta t$ with the transition probabilities $p_{x,y}$ defined in Eq. \eqref{3}. 

For the Model I we find
\begin{align}
\theta_{j \geq 0} = \frac{\phi_0^2}{\phi_j^2} \,, \quad \eta_{j \geq 1} = \frac{H_{0, j}}{W_0 \phi^2_j}  =  \frac{1}{W_0 \phi^2_j} \sum_{x=0}^{j-1} \phi_x \phi_{x+1} \,, \eta_0 = 0 \,,
\end{align}
such that the realization-dependent mean first-passage time is given by
\begin{align}
T^{(I)}_N  = \frac{1}{W_0} \sum_{j=0}^{N-1} \frac{1}{\phi_j^2} \sum_{x=j}^{N-1} \phi_x \phi_{x +1} \,.
\end{align}
Correspondingly, the first two moments of $T_N^{(I)}$ are given by
\begin{align}
\begin{split}
&\Big \langle T^{(I)}_N \Big \rangle  = \frac{N^2}{2 W_0} \left \langle \frac{1}{\phi^2} \right \rangle \left \langle \phi \right \rangle^2 + O(N) =  \frac{N^2}{2 W_0} e^{3 \beta^2 \sigma^2/4} + O(N) \,, \\
&\Big \langle \left(T^{(I)}_N\right)^2 \Big \rangle  = \frac{N^4}{4 W_0^2} \left \langle \frac{1}{\phi^2} \right \rangle^2 \left \langle \phi \right \rangle^4 + O\left(N^3\right) =  \frac{N^4}{4 W_0^2} e^{3 \beta^2 \sigma^2/2} + O\left(N^3\right) \,.
\end{split}
\end{align}
Turning to the Model II and noting that the transition probabilities depend on the on-site potentials at both neighboring sites, we have to define $U_{x=-1}$ and $U_{x=N+1}$. For simplicity, we suppose that we have impermeable barriers at these two sites such that $U_{x=-1} = U_{x=N+1} = \infty$ and hence, $\phi_{-1} = \phi_{N+1} = \infty$. Then, after some algebra we have  
\begin{align}
\begin{split}
\eta_0 = 0 \,, \quad \eta_{j  \geq 1} &= \delta t \left(\frac{1}{\phi_{j-1} \phi_j} + \frac{1}{\phi_{j} \phi_{j+1}}\right) H_{0,j} = \delta t \left(\frac{1}{\phi_{j-1} \phi_j} + \frac{1}{\phi_{j} \phi_{j+1}}\right) \sum_{x=0}^{j-1} \phi_x \phi_{x+1} \,, \\
\theta_{j} &= \phi_0 \phi_1 \left(\frac{1}{\phi_{j-1} \phi_j} + \frac{1}{\phi_{j} \phi_{j+1}}\right) \,.
\end{split}
\end{align}
Inserting the above equations into Eq. \eqref{TN}, we get the following expression for the realization-dependent mean first-passage time
\begin{align}
\label{TNII}
T^{(II)}_N = \delta t \sum_{j=0}^{N-1} \left(\frac{1}{\phi_{j-1} \phi_j} + \frac{1}{\phi_{j} \phi_{j+1}}\right)  \sum_{x=j}^{N - 1} \phi_x \phi_{x+1} \,.
\end{align}
Averaging the expression \eqref{TNII} we find
\begin{align}
\begin{split}
&\Big \langle T^{(II)}_N \Big \rangle  = \frac{(2 \delta t) N^2}{2} \left \langle \frac{1}{\phi} \right \rangle^2 \left \langle \phi \right \rangle^2 + O(N) =  \frac{(2 \delta t) N^2}{2} e^{\beta^2 \sigma^2/2} + O(N) \,, \\
&\Big \langle \left(T^{(II)}_N\right)^2 \Big \rangle  = \frac{(2 \delta t)^2 N^4}{4 W_0^2} \left \langle \frac{1}{\phi} \right \rangle^4 \left \langle \phi \right \rangle^4 + O\left(N^3\right) =  \frac{(2 \delta t)^2 N^4}{4} e^{\beta^2 \sigma^2} + O\left(N^3\right) \,.
\end{split}
\end{align}
Lastly, we turn to the Model III for which the rates obey Eq. \eqref{rates3} and hence,
\begin{align}
\begin{split}
\theta_{j \geq 0} &= \frac{\phi_0^2}{\phi_j^2}  \,,\quad
\eta_{j \geq 0}  = \frac{j}{W_0 \phi_j^2} \,,
\end{split}
\end{align}    
such that the realization-dependent mean first-passage time attains a very compact form
\begin{align}
T_N^{(III)}  = \frac{1}{W_0} \sum_{j=0}^{N-1} \frac{(N - j)}{\phi_j^2} \,.
\end{align}
Respectively, the first and second moments of $T_N^{(III)}$ evaluated by averaging the above expression 
with the probability density function in Eq. \eqref{Pphim} are given by 
\begin{align}
\begin{split}
&\Big \langle T^{(III)}_N \Big \rangle  = \frac{N (N + 1)}{2 W_0} e^{\beta^2 \sigma^2/2} \left(1 + \erf\left(\beta \sigma/\sqrt{2}\right)\right) \,, \\
&\Big \langle \left(T^{(III)}_N\right)^2 \Big \rangle  = \frac{(N-1) N (N+1) (3 N+2)}{12 W_0^2} e^{\beta^2 \sigma^2} \left(1 + \erf\left(\beta \sigma/\sqrt{2}\right)\right)^2 \\
&+ \frac{N (2 N^2 + 3 N + 1)}{6 W_0^2} e^{2 \beta^2 \sigma^2} \left(1 + \erf\left(\sqrt{2} \, \beta \sigma\right)\right) \,.
\end{split}
\end{align}
It is then straightforward to conclude that for  all three models the relative variance of the realization-dependent mean first-passage times
obeys
\begin{align}
	\label{RT}
R_T = O(1/N) \,,
\end{align}
which means that these properties are self-averaging in the limit $N \to \infty$. However, as we have already remarked, 
this just shows the trend that fluctuations become relatively less important the farther the target is located away from the starting point. For all practically important applications $N$ is essentially finite and fluctuations are therefore important. 

\begin{figure}[h!]
	\centering
	\includegraphics[width=160mm,angle=0]{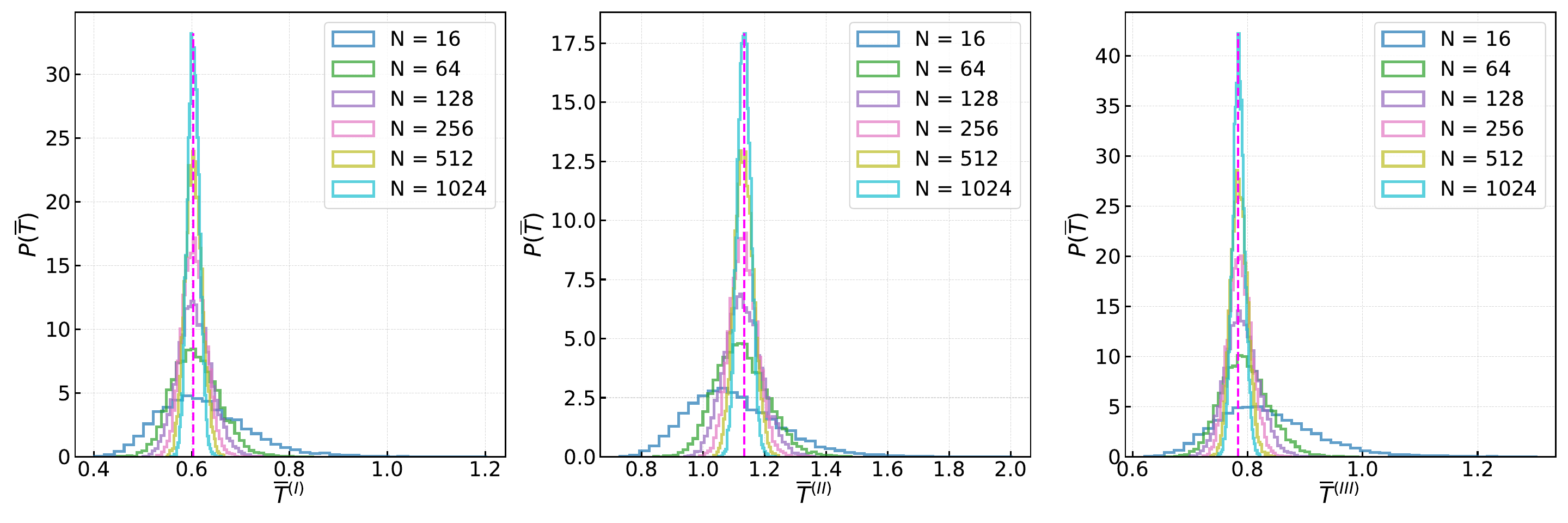}  
	\caption{Numerically-evaluated normalized probablity density functions of the reduced mean first-passage times $\overline{T} = T_N/N^2$ for the Models I (left panel), II (middle panel) and III (right panel) and different values of the distance $N$ to the target (see the insets). Vertical dashed lines indicate positions the disorder-averaged mean first passage times. }
	\label{fig:5} 
\end{figure}

In Fig. \ref{fig:5} we present numerically-evaluated probability density functions of the realization-dependent reduced mean first-passage times $\overline{T} = T_N/N^2$ for several values of the distance $N$ to the target and three models under study. We observe that, in full accord with our prediction in Eq. \eqref{RT}, the probability density functions become narrower and more sharply peaked at the disorder-averaged values of the mean first-passage times, the larger $N$ is.  

\section{The  realization-dependent diffusion coefficient}
\label{sec:6}

Most of existing analyses of the effective diffusion coefficient of particles diffusing in a random Gaussian potential focus on one-dimensional, periodic, \textit{continuous-space} systems with period $L$ (see, e.g., \cite{dean3} and \cite{dean}). These studies often build upon the well-known expression for the effective diffusion coefficient in a periodic continuum with an arbitrary potential $U_x$, originally derived by Lifson and Jackson \cite{LJ} (see also \cite{festa})
\begin{align}
\label{LJ}
D_L = \frac{D_0 L^2}{\int^L_0 dx \, e^{\beta U_x} \int^L_0 dy \, e^{-\beta U_y}} \,,
\end{align}
where $D_0$ is the diffusion coefficient in absence of the external potential $U_x$. Averaging the above expression
and going to the limit $L \to \infty$, in order to describe the behavior in infinitely large systems, 
several notable results have been obtained. In particular, it was shown that for a Gaussian random potential with the probability density function in Eq. \eqref{Gauss} one has
\cite{zwan} (see also \cite{dean} for a review)
\begin{align}
\label{Zwanzig}
\Big \langle D_{L} \Big \rangle =  D_0 \, e^{- \beta^2 \sigma^2} \,.
\end{align}
This result demonstrates that the diffusion coefficient decreases in the presence of a Gaussian random potential -- a physically intuitive outcome. Remarkably, the reduction factor takes the form of a simple exponential function of 
$\beta^2 \sigma^2$, exhibiting therefore a pronounced super-Arrhenius dependence on temperature.

In this Section we seek for the analogues of 
 the result in Eq. \eqref{LJ} for  the random walks on \textit{discrete} periodic chains with $N$ sites and also for different types of dynamics, as defined in our Models I, II and III. To this end, we will make use of Derrida’s general, though rather intricate, expression for the effective diffusion coefficient 
 $D_N$, valid for an arbitrary set of transition rates \cite{derdiff}. In \ref{C} we present details of the derivation of  $D_N$ for the three models under study.
 Then, we will determine the moments of  
 $D_N$, discuss their dependence on $\beta \sigma$ and   
on this basis will demonstrate that the realization-dependent diffusion coefficient is self-averaging in the limit $N \to \infty$, regardless of the dynamical scenario.

\subsection{Model I: Random force-like model}

In \ref{C} we show that for such  a model the Derrida's formula \cite{derdiff} reduces to the following expression
\begin{align}
\label{DRF}
D_N = \frac{W_0 N^2}{\sum_{x=1}^N e^{- \beta U_x} \sum_{y=1}^N e^{\frac{\beta}{2} (U_y + U_{y+1})}}  =  \frac{W_0 N^2}{\sum_{x=1}^N \phi_x^{-2} \sum_{y=1}^N \phi_y \phi_{y+1}} \,, U_{N+x} = U_x \,.
\end{align}
The above 
expression does not have an apparent symmetry $U_x \to - U_x$, unlike the original Lifson-Jackson formula \eqref{LJ}, but converges to it if we rewrite Eq. \eqref{DRF} for an arbitrary inter-site spacing $a$ and then turn to the continuous-space limit $a \to 0$.

We introduce next an auxiliary function
\begin{align}
\Theta^{(I)}_N(\lambda_1,\lambda_2) = \left \langle \exp\left(  - \Psi^{(I)}_N(\lambda_1,\lambda_2) \right) \right \rangle \,,
\end{align}
where
\begin{align}
\Psi^{(I)}_N(\lambda_1,\lambda_2) = \frac{\lambda_1}{N} \sum_{x=1}^N \phi_x^{-2} + \frac{\lambda_2}{N}  \sum_{x=1}^N \phi_x \phi_{x+1} \,.
\end{align}
One may readily notice that the moments of the realization-dependent diffusion coefficient in Eq. \eqref{DRF} can be evaluated by a mere integration of  $\Theta^{(I)}_N(\lambda_1,\lambda_2)$,
\begin{align}
\Big \langle D_N^q \Big \rangle = \frac{W^q_0}{\Gamma^2(q)} \int^{\infty}_0 \int^{\infty}_0 d\lambda_1 \, d\lambda_2 (\lambda_1 \lambda_2)^{q-1}   \, \Theta^{(I)}_N(\lambda_1,\lambda_2)  \,.
\end{align}

We turn to the limit $N \to \infty$ and consider the first two moments of the functional $\Psi^{(I)}_N(\lambda_1,\lambda_2)$. These moments can be evaluated very straightforwardly to give
\begin{align}
\begin{split}
\Big \langle \Psi^{(I)}_N(\lambda_1,\lambda_2) \Big \rangle & = \lambda_1 \left \langle \frac{1}{\phi^2} \right \rangle +  \lambda_2 \left \langle \phi \right \rangle^2 \\
\Big \langle \left(\Psi^{(I)}_N(\lambda_1,\lambda_2)\right)^2 \Big \rangle & = \lambda_1^2  \left \langle \frac{1}{\phi^2} \right \rangle^2 + \lambda_2^2 \left \langle \phi \right \rangle^4 + 2 \lambda_1 \lambda_2  \left \langle \frac{1}{\phi^2} \right \rangle \left \langle \phi \right \rangle^2 + O(1/N) \,.
\end{split}
\end{align}
Since the leading terms in the second equation can be expressed as the square of the right-hand-side of the first equation, it follows that the function 
 $\Psi^{(I)}_N(\lambda_1,\lambda_2)$ is self-averaging in the limit $N \to \infty$, such that
\begin{align}
\Theta^{(I)}_{N  \to \infty}(\lambda_1,\lambda_2) = \exp\left(  -  \lambda_1 \left \langle \frac{1}{\phi^2} \right \rangle -  \lambda_2 \left \langle \phi \right \rangle^2 \right) \,.
\end{align}
Consequently, the moments of $D_N$ in an infinite system obey
\begin{align}
\label{momDRFT}
\Big \langle D_{N\to \infty}^q \Big \rangle = \frac{W^q_0}{\left \langle 1/\phi^2 \right \rangle^q \left \langle \phi \right \rangle^{2 q}} = W_0^q  \exp\left(- 3 \, q \, \beta^2 \sigma^2/4\right) \,.
\end{align}
The effective diffusion coefficient is therefore self-averaging in the limit $N \to \infty$. 

\subsection{Model II:  Random walks with randomized stepping-times}

In \ref{C} we show that for this model the realization-dependent diffusion coefficient obeys the following symmetric form
\begin{align}
\label{DDD}
\begin{split}
D_N &= \dfrac{N^2}{2 \delta t \left(\sum_{x=1}^N e^{-\frac{\beta}{2}\left(U_x + U_{x+1}\right)}\right)\left(\sum_{y=1}^N e^{\frac{\beta}{2}\left(U_y + U_{y+1}\right)}\right)} \\&= \dfrac{N^2}{2 \delta t \sum_{x=1}^N \dfrac{1}{\phi_x \phi_{x+1}} \sum_{y=1}^N \phi_y \phi_{y+1}} \,, U_{x} = U_{x+N} \,, \\
\end{split}
\end{align}
which resembles the Lifson-Jackson formula but with a rather non-evident "discretization" scheme. This compact form preserves the symmetry $U_x \to - U_x$ 
and reduces to the expression \eqref{LJ} upon the passage to the continuous-space limit.

As above, we introduce an auxiliary function
\begin{align}
\Theta^{(II)}_N(\lambda_1,\lambda_2) = \left \langle \exp\left(  - \Psi^{(II)}_N(\lambda_1,\lambda_2) \right) \right \rangle \,,
\end{align}
where $\Psi^{(II)}_N(\lambda_1,\lambda_2)$ is now given by
\begin{align}
\Psi^{(I)}_N(\lambda_1,\lambda_2) = \frac{\lambda_1}{N} \sum_{x=1}^N \frac{1}{\phi_x \phi_{x+1}} + \frac{\lambda_2}{N}  \sum_{x=1}^N \phi_x \phi_{x+1} \,.
\end{align}
Once this function is determined, the moments of $D_N$ of an arbitrary (not necessarily integer) order $q$ are then simply given by the integral
\begin{equation}
\Big \langle  D_N^q \Big \rangle = \frac{1}{\left(2 \delta t\right)^q \Gamma^2(q)} \int^{\infty}_0 \int^{\infty}_0 d\lambda_1 d\lambda_2 (\lambda_1 \lambda_2)^{q-1} \Theta^{(II)}_N(\lambda_1,\lambda_2) \,.
\end{equation}
Further on, we consider the first two moments of the functional $\Psi^{(II)}_N(\lambda_1,\lambda_2)$ to get
\begin{align}
\begin{split}
\Big \langle \Psi^{(II)}_N(\lambda_1,\lambda_2) \Big \rangle & = \lambda_1 \left \langle \frac{1}{\phi} \right \rangle^2 +  \lambda_2 \left \langle \phi \right \rangle^2 \\
\Big \langle \left(\Psi^{(II)}_N(\lambda_1,\lambda_2)\right)^2 \Big \rangle & = \lambda_1^2  \left \langle \frac{1}{\phi} \right \rangle^4 + \lambda_2^2 \left \langle \phi \right \rangle^4 + 2 \lambda_1 \lambda_2  \left \langle \frac{1}{\phi} \right \rangle^2 \left \langle \phi \right \rangle^2 + O(1/N) \,.
\end{split}
\end{align}
Again, we observe that the leading terms in the equation in the second line are just the full square of the right-hand-side of the equation in the first line, which signifies
that the functional $\Psi^{(II)}_N(\lambda_1,\lambda_2)$ is self-averaging in the limit $N \to \infty$. As a consequence,
\begin{align}
\Theta^{(I)}_{N  \to \infty}(\lambda_1,\lambda_2) = \exp\left(  -  \lambda_1 \left \langle \frac{1}{\phi} \right \rangle^2 -  \lambda_2 \left \langle \phi \right \rangle^2 \right) \,, 
\end{align}
such that the moments of the effective diffusion coefficient in an infinite system read
\begin{align}
\label{momDKAW}
\Big \langle D_{N\to \infty}^q \Big \rangle = \frac{1}{(2 \delta t)^q \left \langle 1/\phi \right \rangle^{2 q} \left \langle \phi \right \rangle^{2 q}} = \frac{1}{(2 \delta t)^q}  \exp\left(- q \,\beta^2 \sigma^2/2\right) \,.
\end{align}
The effective diffusion coefficient is evidently self-averaging in the limit $N \to \infty$. 

We close these two subsections with the following two remarks:\\
-- Neither the expression \eqref{momDRFT}
nor the expression \eqref{momDKAW} reproduce \textit{exactly} the result in Eq. \eqref{Zwanzig},  
obtained in \cite{zwan} for the continuous-space expression for the effective diffusion coefficient in Eq. \eqref{LJ}:  both dynamical scenarios yield indeed a simple exponential, super-Arrhenius dependence on $\beta^2 \sigma^2$ but predict somewhat different values of the numerical factors.\\
-- At the same time,  a "direct" discretization of the continuous-space Lifson-Jackson formula \eqref{LJ} gives
\begin{align}
\label{discr}
D_N = \frac{D_0 N^2}{\sum_{x = 1}^N e^{\beta U_x} \sum_{y = 1}^N e^{ - \beta U_y}}  = \frac{D_0 N^2}{\sum_{x = 1}^N \phi_x^2 \sum_{y = 1}^N \phi_y^{-2}} \,.
\end{align}
Using the above arguments about the self-averaging in the limit $N \to \infty$, we find that 
\begin{align}
\Big \langle D^q_{N \to \infty} \Big \rangle = D_0^q \exp\left(- q \beta^2 \sigma^2\right) \,,
\end{align}
which gives precisely the expression \eqref{Zwanzig} for $q  = 1$. On the other hand, we could not find (and doubt if they exist) the transition rates $W_{x, x \pm 1}$ for the microscopic stochastic dynamics on the discrete lattice which will give exactly the form in Eq. \eqref{discr}.

\subsection{Model III: Gaussian random trap model}

We show in \ref{C} that within this dynamical scenario the  realization-dependent effective diffusion coefficient attains a particularly simple form:
\begin{align}
	\label{333}
D_N = \frac{W_0 N}{\sum_{x=1}^N e^{-\beta U_x}} = \frac{W_0 N}{\sum_{x=1}^N \phi_x^{-2}} \,.
\end{align}
As above, our next step consists in introducing an
auxiliary function
\begin{align}
\Theta^{(III)}(\lambda)  = \lim_{N \to \infty} \left \langle \exp\left(- \frac{\lambda}{N } \sum_{x=1}^N \phi_x^{-2}\right) \right \rangle \,,
\end{align}
such that the moments of $D_{N \to \infty}$ can be obtained by performing the integral
\begin{align}
	\label{mom3}
\Big \langle D^q_{N \to \infty}  \Big \rangle = \frac{W_0^q}{\Gamma(q)} \int^{\infty}_0 \lambda^{q - 1} \, d\lambda \, \Theta^{(III)}(\lambda)
\end{align}
The auxiliary function $\Theta^{(III)}(\lambda)$ can be calculated very straightforwardly to give
\begin{align}
\begin{split}
\Theta^{(III)}(\lambda)  &= \lim_{N \to \infty} \left[\int^1_0 d\phi \, e^{- \lambda/N \phi^2}  \, P_{-}(\phi)\right]^N = \lim_{N \to \infty} \left[1 - \int^1_0 d\phi \, \left(1 - e^{- \lambda/N \phi^2}\right)  \, P_{-}(\phi)\right]^N \\
& = \exp\left(- \lambda \int^1_0 \frac{d\phi}{\phi^2} \, P_{-}(\phi)\right)  = \exp\left(- \lambda e^{\beta^2 \sigma^2} \left(1 + \erf\left(\beta \sigma/\sqrt{2}\right) \right)\right)  \,.
\end{split}
\end{align}
Inserting the latter expression into Eq. \eqref{mom3} and performing the integral, we find
\begin{align}
	\label{diffIII}
\Big \langle D^q_{N \to \infty}  \Big \rangle = \frac{W_0^q}{\left(1 + \erf\left(\beta \sigma/\sqrt{2}\right)\right)^q} e^{-q \beta^2 \sigma^2/2} \,.
\end{align}
One can readily check directly that 
 also for the Model III the diffusion coefficient is self-averaging in the limit $N \to \infty$.

\begin{figure}[h!]
	\centering
	\includegraphics[width=160mm,angle=0]{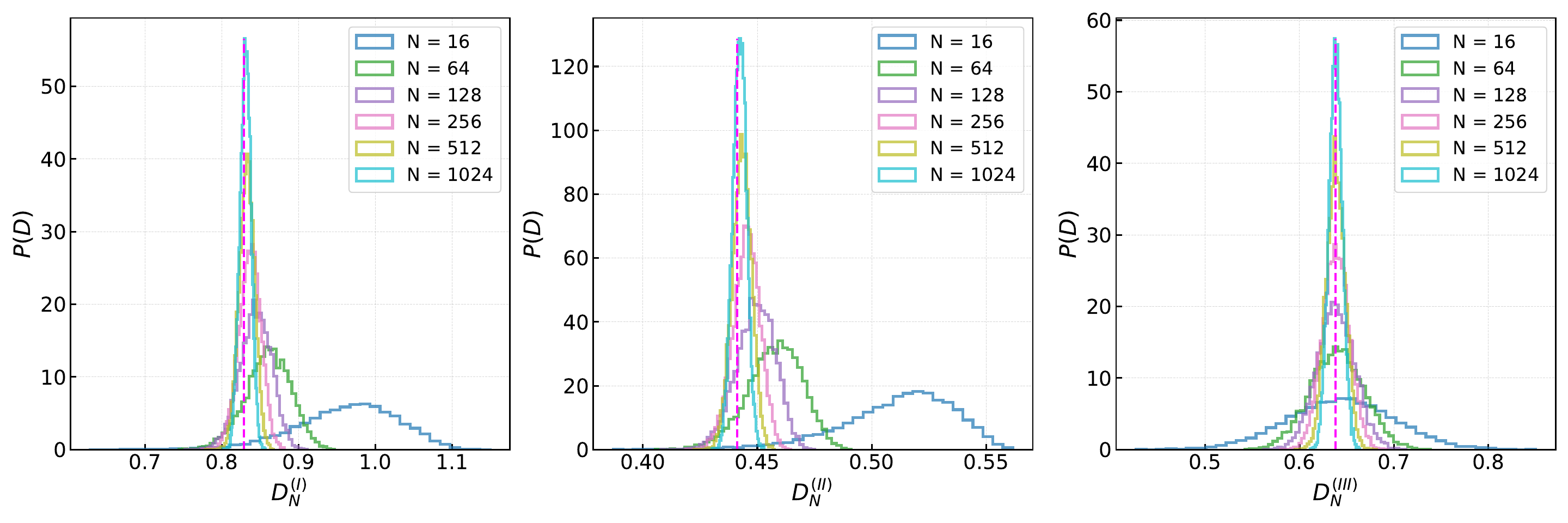}  
	\caption{Numerically-evaluated normalized probability density functions of the effective diffusion coefficients for the Models I (left panel), II (middle panel) and III (right panel) for different values of $N$ ranging from $N = 16$ to $N = 1024$ (see the insets). Vertical dashed lines indicate the limiting ($N \to \infty$) values of the disorder-averaged diffusion coefficients (see Eqs. \eqref{momDRFT}, \eqref{momDKAW} and \eqref{diffIII} with $q = 1$).
		}
	\label{fig:7} 
\end{figure}

Finally, for all three models, we numerically compute the probability density functions of the diffusion coefficients for several values of 
$N$. As shown in Fig. \ref{fig:7}, the distributions become progressively narrower and increasingly concentrated around the disorder-averaged diffusion coefficients (indicated by the vertical dashed lines) as $N$ grows.

\section{Conclusions}
\label{sec:7}

In conclusion, we have revisited the long-standing problem of random walks in the presence of a quenched Gaussian random potential, a topic that has attracted considerable attention in the past. Specifically, we analyzed the continuous-time dynamics of a particle performing nearest-neighbor random walks on a one-dimensional lattice, where each site 
$x$ hosts  a quenched random potential 
$U_x$ that governs the local transition rates. The onsite potentials 
$U_x$
are assumed to be independent and identically distributed Gaussian random variables. 
 
We have considered three distinct dynamical scenarios, each characterized by a different dependence of the transition rates on the onsite potentials $U_x$: (i) a random-force-like model, (ii) a continuous-time random walk with exponentially distributed waiting times, where the transition probabilities depend on the potentials on neighboring sites, and (iii) a Gaussian random trap model.

For all three scenarios, we analyzed the statistical properties of five realization-dependent random variables: (a) the steady-state probability current through a finite chain with  
$N$
 sites, (b) its reciprocal, representing the resistance of a finite interval with  
$N$
sites with respect to random transport, (c) the splitting probability -- the probability that a particle starting at site 
$x_0$
within the interval 
$(0,N)$
first reaches the left boundary without ever visiting the right boundary, (d) the mean first-passage time across a finite interval of length 
$N$, and (e) the diffusion coefficient in a finite periodic chain with 
$N$
 sites.

For all these random variables we have
calculated analytically the leading (in the large-$N$ limit) asymptotic behavior  of their moments determining the dependence on the strength of disorder and the temperature. We have shown that these moments exhibit an interesting super-Arrhenius dependence on the temperature, which is also quite different depending on the dynamical scenario.      
Comparing the averaged and the typical behavior we have demonstrated that the moments of the resistance and currents  are supported by atypical realizations of disorder. 

For intermediate values of $N$, we have performed numerical analyses, employing exact numerical averaging of the formal expressions that define these properties for a given sequence of $U_x$ values. 

Further on, we have addressed a  conceptually important question whether these transport properties exhibit self-averaging with respect to different realizations of disorder in the limit $N \to \infty$. Capitalizing on the derived exact expressions for the moments, we have studied the asymptotic behavior of their relative variances. We have shown that the probability currents through finite chains and the  resistances of these chains are not self-averaging in the limit $N \to \infty$. Interestingly enough, such a lack of self-averaging appears to be a boundary effect; namely, it stems from fluctuations of one (or a few) sites close to the boundary, while the contributions coming from the sites in the bulk of chains are self-averaging. 

Concurrently, we have shown that the realization-dependent 
splitting probability and the realization-dependent
mean first-passage time to a target placed at distance $N$ away from the starting point are \textit{self-averaging} in the limit $N \to \infty$. Clearly, this result just indicates the trend in the behavior of fluctuations that they become progressively less important the farther is the target away from the starting point. However, for most of practical applications $N$ is essentially finite and our numerical analysis shows that for such $N$ fluctuations are quite significant. 

Lastly, we have shown that for all three dynamical scenarios the realization-dependent diffusion coefficient in periodic chains with $N$ sites is self-averaging in the limit $N \to \infty$. This result is very meaningful because it applies to the dynamics in infinite systems. For finite $N$ the diffusion coefficient also exhibits essential sample-to-sample fluctuations.

Summing up, our results demonstrate that even in the relatively simple case of Gaussian disorder, the latter plays a significant role and gives rise to a variety of interesting effects. One can anticipate an even more pronounced impact of disorder on the transport properties in models such as the random trap model with an exponential distribution of traps' depths or the Touya-Dean model \cite{touya}, where not all moments of the relevant random, realization-dependent variables exist. We plan to study these cases in our future work.

\section*{Acknowledgments}

The authors wish to thank E. Agliari, T. Akimoto, D. S. Dean, E. Barkai and L. A.  Pastur for helpful discussions. This project has been supported by funding from the 2021 first FIS (Fondo Italiano per la
Scienza) funding scheme (FIS783—SMaC—Statistical Mechanics and Complexity University and Research)
from Italian MUR (Ministry of University and Research).

\appendix

\section{Realization-dependent probability current}
\label{A}

Consider a finite chain in which we fix the value of the probability $P_{x=0} = P(x=0,t) = c_0$ (see Eq. \eqref{ME}) 
at the site $x=0$, and place a perfect sink at the site $x = N$, such that  $P_{x=N} = P(x=N,t) = 0$. 
Then, in such a situation the evolution equations \eqref{ME} become 
\begin{align}
\begin{split}
& \dot{P}_1=  - \left(W_{1,0} + W_{1,2}\right) P_1 + W_{0,1} c_0  + W_{2,1} P_2 \,, \\
&\dot{P}_2 = - \left(W_{2,1} + W_{2,3}\right) P_2 + W_{1,2} P_1  + W_{3,2} P_3 \,, \\
&\dot{P}_3 = - \left(W_{3,2} + W_{3,4}\right) P_3 + W_{2,3} P_2  + W_{4,3} P_4 \,, \\
&\ldots \\
&\dot{P}_{N-1} = -  \left(W_{N-1,N-2} + W_{N-1,N}\right) P_{N-1} + W_{N-2,N-1} P_{N-2} \,.
\end{split}
 \end{align}
Setting next $\dot{P}_x = 0$ for all $x$,  and solving the resulting linear system of equations recursively, we get for $P_{N-1}$:
\begin{align}
P_{N-1} = \frac{W_{0,1} c_0}{W_{N-1,N} \, \Sigma_N} \,,
\end{align}
where $\Sigma_N$ obeys 
\begin{align}
\label{sigma}
\begin{split}
&\Sigma_N = 1 + \frac{W_{1,0}}{W_{1,2}} +  \frac{W_{1,0} W_{2,1}}{W_{1,2} W_{2,3}} + \frac{W_{1,0} W_{2,1} W_{3,2}}{W_{1,2} W_{2,3} W_{3,4}} + \frac{W_{1,0} W_{2,1} W_{3,2} W_{4,3}}{W_{1,2} W_{2,3} W_{3,4} W_{4,5}} + \nonumber\\
& \ldots + \frac{W_{1,0}  W_{2,1} W_{3,2} W_{4,3} \ldots W_{N-2,N-3}}{W_{1,2} W_{2,3} W_{3,4} W_{4,5} \ldots W_{N-2,N-1 }} + \frac{W_{1,0} W_{2,1} W_{3,2} W_{4,3} \ldots W_{N-2,N-3} W_{N-1,N-2}}{W_{1,2} W_{2,3} W_{3,4} W_{4,5} \ldots W_{N-2,N-1} W_{N-1,N}}  \,.
\end{split}
\end{align}
Respectively, the realization-dependent current to the sink at $x = N$ obeys
\begin{align}
j_N = W_{N-1,N} P_{N-1} = \frac{W_{0,1} c_0}{\Sigma_N} \,.
\end{align}
For the Model I the rates are defined in Eqs. \eqref{rates}, which gives  
\begin{align}
\label{sigma1}
&\Sigma_N^{(I)} = \frac{1}{\phi_0 \phi_1} \left(\phi_0 \phi_1 + \phi_1 \phi_2 + \phi_2 \phi_3 + \ldots + \phi_{N-1} \phi_N\right) \,.
\end{align}
Hence, the realization-dependent current to the sink is given by
\begin{align}
j_N^{(I)} = \frac{\phi_0 W_0 c_0}{\phi_1 \Sigma_N^{(I)}} \,,
\end{align}
from which we find the first expression 
in Eq. \eqref{jN}.

For the Model II, the rates obey
\begin{align}
W_{x,x \pm 1} = \frac{p_{x, x \pm 1}}{\delta t} \,,
\end{align}
where the normalized transition probabilities are defined in Eq. \eqref{33}. Upon some algebra we find that 
$\Sigma_N$ in Eq. \eqref{sigma} for this model is exactly the same as for the Model I, i.e., it
obeys Eq. \eqref{sigma1}. Consequently, the steady-state realization-dependent current to the sink follows
\begin{align}
j_N^{(II)} = \frac{\phi_{-1}}{(\phi_{-1} + \phi_1)}  \frac{c_0}{\delta t  \Sigma_N^{(I)}} \,.
\end{align}
Supposing, for simplicity, that the potential on the site $x = - 1$ is exactly the same as at the site $x=1$, 
we get the second expression in Eq. \eqref{jN}.

For the Model III with the rates given by Eq. \eqref{rates3}, one finds that $\Sigma_N$ is independent of the depths of the traps and is simply given by
\begin{align}
\Sigma_N^{(III)} = N \,,
\end{align}
i.e., is not fluctuating. 
Correspondingly, the steady-state  current obeys
\begin{align}
j_N^{(III)} = \frac{W_0 c_0 \phi_0^2}{N} \,,
\end{align}
from which we read of the third expression in Eq. \eqref{jN}.

\section{Realization-dependent splitting probability}
\label{B}
In this Appendix we derive an explicit expression for the splitting probability $E_-$. We consider a continuous-time random walk on a finite one-dimensional lattice with sites $x = 0,\,1,\,\dots,\,N$, with the transition rates  $W_{x,\,x+1}$ and $W_{x,\,x-1}$ for jumps from the site $x$ to the neighboring sites $x-1$ and $x+1$.
Let $E_-(x_0)$ denote the probability that, starting from site $x_0$, the random walk
visits the left boundary at $x=0$ before reaching the right boundary at $x=N$.

As shown in \cite{vanKampen1992} the splitting probability $E_-(x)$ (here, the walk is assumed to start at arbitrary point $x$, excluding 
$x=0$ and $x=N$) satisfies the \emph{backward} master equation of the form
\begin{equation}
	\label{eq:backwsplit}
	W_{x,x+1} \big[ E_-(x+1) - E_-(x) \big]
	+ W_{x,x-1} \big[ E_-(x-1) - E_-(x) \big] = 0, \quad x=1,\dots,N-1 \,,
\end{equation}
which is to be solved subject to trivial boundary conditions :
\begin{equation}
    E_-(x=0) = 1, 
    \qquad 
    E_-(x=N) = 0.
\end{equation}

Equation~\eqref{eq:backwsplit} can be conveniently rewritten in terms of finite differences $\Delta E_-(x)=E_-(x+1)-E_-(x)$,
\begin{equation}
    \Delta E_-(x) = \frac{W_{x,\,x-1}}{W_{x,\,x+1}}\, \Delta E_-(x-1),\qquad 1\leq x< N \,.
\end{equation}
and solved recursively to give
\begin{equation}
    \Delta E_-(x) = \Delta E_-(0) \prod_{k=1}^{x} \frac{W_{k,\,k-1}}{W_{k,\,k+1}} \,.
\end{equation}
The above expression yields the following explicit expression in terms of $E_-(x)$:
\begin{equation}
    E_-(x) = 1 + \Delta E_-(0) \Sigma_x \,,
\end{equation}
in which $\Delta E_-(0)$ is as yet an undefined parameter and
the sum $\Sigma_x$ has been introduced in \ref{A},
\begin{align}
\Sigma_x &= \displaystyle 1 + \sum_{j=1}^{x-1} \prod_{k=1}^{j} \frac{W_{k,\,k-1}}{W_{k,\,k+1}}.
\end{align}
Lastly, the boundary condition \( E_-(N) = 0 \) ensures that
\[
\Delta E_-(0) = -\frac{1}{\Sigma_N},
\]
and hence, the general solution  for $x=x_0$ reads
\begin{equation}
\label{eq:solsplit}
    E_-(x_0) = 1-\frac{\displaystyle \Sigma_{x_0}}{\displaystyle \Sigma_N}=
    \frac{
        \displaystyle \sum_{j=x_0}^{N-1} 
        \prod_{k=1}^{j} 
        \frac{W_{k,\,k-1}}{W_{k,\,k+1}}
    }{
        1 + \displaystyle \sum_{j=1}^{N-1} 
        \prod_{k=1}^{j} 
        \frac{W_{k,\,k-1}}{W_{k,\,k+1}}
    }.
\end{equation}
We note next that the above expression~\eqref{eq:solsplit} can be cast into a compact form in Eq. \eqref{eq:sol_bella}, 
where \( \tau_{x_0}^- \) and \( \tau_{N-x_0}^+ \) represent, respectively, 
the resistances of the intervals $(0,x_0-1)$ and $(x_0+1,N)$ from the left and from the right of the starting point $x_0$:
\begin{equation}
\label{eq:res1res2}
\begin{split}
   \displaystyle\tau_{x_0}^- &= \frac{1}{W_{x_0,\,x_0-1}}
   \left( 1 + \frac{W_{x_0-1,\,x_0}}{W_{x_0-1,\,x_0-2}}
   +\dots+ \frac{W_{x_0-1,\,x_0} \cdots W_{1,\,2}}
          {W_{x_0-1,\,x_0-2} \cdots W_{1,\,0}} \right),\\
   \displaystyle\tau_{N-x_0}^+ &= \frac{1}{W_{x_0,\,x_0+1}}
   \left( 1 + \frac{W_{x_0+1,\,x_0}}{W_{x_0+1,\,x_0+2}}
   +\dots+ \frac{W_{x_0+1,\,x_0} \cdots W_{N-1,\,N-2}}
          {W_{x_0+1,\,x_0+2} \cdots W_{N-1,\,N}} \right).   
\end{split}
\end{equation}
This is done by simply noticing that Eq. \eqref{eq:solsplit}
can be formally rewritten as
\begin{equation}
    \begin{split}
        E_-(x_0)&=\frac{
        \displaystyle \sum_{j=x_0}^{N-1} 
        \prod_{k=1}^{j} 
        \frac{W_{k,\,k-1}}{W_{k,\,k+1}}
    }{1+
        \displaystyle \sum_{j=1}^{N-1} 
        \prod_{k=1}^{j} 
        \frac{W_{k,\,k-1}}{W_{k,\,k+1}}
    }=\\ 
    &=\frac{\displaystyle
        \frac{W_{1,\,0}\dots W_{x_0\,x_0-1}}{W_{1,\,2}\dots W_{x_0,\,x_0+1}}+\dots+\frac{W_{1,\,0}\dots W_{N-1,\,N-2}}{W_{1,\,2}\dots W_{N-1,\,N}}
    }{ \displaystyle
    1+\frac{W_{1,\,0}}{W_{1,\,2}}+\dots+\frac{W_{1,\,0}\dots W_{x_0,\,x_0-1}}{W_{1,\,2}\dots W_{x_0,\,x_0+1}}+\dots+\frac{W_{1,\,0}\dots W_{N-1,\,N-2}}{W_{1,\,2}\dots W_{N-1,\,N}}
    }\\ \\
    &=
\left(
1
+\frac{W_{x_0+1,\,x_0}}{W_{x_0+1,\,x_0+2}}
+\dots+
\frac{W_{x_0+1,\,x_0}\dots W_{N-1,\,N-2}}
     {W_{x_0+1,\,x_0+2}\dots W_{N-1,\,N}}
\right)\quad \bigg /\\
&\begin{split}\Bigg(
\frac{W_{1,\,2}\dots W_{x_0,\,x_0+1}}
     {W_{1,\,0}\dots W_{x_0,\,x_0-1}}
+\dots+
\frac{W_{x_0,\,x_0+1}}
     {W_{x_0,\,x_0-1}}
+1
+\frac{W_{x_0+1,\,x_0}}{W_{x_0+1,\,x_0+2}}
+\\ \dots+
\frac{W_{x_0+1,\,x_0}\dots W_{N-1,\,N-2}}
     {W_{x_0+1,\,x_0+2}\dots W_{N-1,\,N}}\Bigg)
\end{split}
    \end{split}
\end{equation}
and then  factoring out the common product
$W_{1,\,0}\cdots W_{x_0,\,x_0-1}/W_{1,\,2}\cdots W_{x_0,\,x_0+1}$
from both numerator and denominator.


Consider next the model-specific forms of the splitting probability $E_-(x_0)$.

Model I. 
Using the general expressions derived in \ref{A}, we have that the resistances of the intervals from the left and
from the right from the starting site $x_0$ attain the following forms:
\begin{equation}
	\begin{split}
	\tau_{x_0}^- &=
	\frac{1}{\phi^2_x}
	\left(
	\phi_{x_0}\phi_{x_0-1}+\phi_{x_0-1}\phi_{x_0-2}+\dots+\phi_{1}\phi_{0}
	\right), \\[2mm]
	\tau_{N-x_0}^+ &=
	\frac{1}{\phi^2_{x_0}}
	\left(
	\phi_{x_0}\phi_{x_0+1}+\phi_{x_0+1}\phi_{x_0+2}+\dots+\phi_{N-1}\phi_{N}
	\right) \,,
    \end{split}
\end{equation}
where $\phi_x$ is defined in the main text.
This gives for the Model I the following expression
\begin{align}
\label{z0}
E_-^{(I)}(x_0)=\frac{\phi_{x_0}\phi_{x_0+1}+\phi_{x_0+1}\phi_{x_0+2}+\dots+\phi_{N-1}\phi_{N}}{\phi_{0}\phi_{1}+\dots+\phi_{N-1}\phi_{N}}=\frac{H_{x_0,\,N}}{H_{0,N}} \,.
\end{align}

Model II. 
In this case the resistances admit the following form
\begin{equation}
	\begin{split}
	\tau_{x_0}^- &=
	\frac{\phi_{x_0+1}+\phi_{x_0-1}}{\phi_{x_0+1}\phi_{x_0}\phi_{x_0-1}}
	\left(
	\phi_{x_0}\phi_{x_0-1}+\phi_{x_0-1}\phi_{x_0-2}+\dots+\phi_{1}\phi_{0}
	\right), \\[2mm]
	\tau_{N-x_0}^+ &=
	\frac{\phi_{x_0-1}+\phi_{x_0+1}}{\phi_{x_0-1}\phi_{x_0}\phi_{x_0+1}}
	\left(
	\phi_{x_0}\phi_{x_0+1}+\phi_{x_0+1}\phi_{x_0+2}+\dots+\phi_{N-1}\phi_{N}
	\right).
    \end{split}
\end{equation}
\noindent
Noticing that the multipliers $(\phi_{x_0-1}+\phi_{x_0+1})/\phi_{x_0-1}\phi_{x_0}\phi_{x_0+1}$ cancel each other, we find that also for the Model II the splitting probability $E_-^{(II)}(x_0)$ obeys Eq. \eqref{z0}. 

Model III. 
For the random trap model the transition rates  are symmetric and hence, 
\begin{align}
	\label{eq:model_res_left}
	\tau_{x_0}^- =
	\frac{x_0}{\phi_{x_0}^2}
	\,, \quad
	\tau_{N-x_0}^+ =
	\frac{N-x_0}{\phi_{x_0}^2}
	.
\end{align}
Consequently, we  find that $E_-^{(III)}(x_0)$ is independent of disorder (i.e., it is not a fluctuating property) and obeys
\begin{align}
	E_-^{(III)}(x_0)=1-\frac{x_0}{N} \,.
\end{align}

\section{Realization-dependent diffusion coefficient}
\label{C}

In this appendix, we present the derivation of our expressions 
for the realization-dependent diffusion coefficients in the Models I, II and III (see Eqs. \eqref{DRF}, \eqref{DDD} and \eqref{333}) taking advantage  of the general result due to Derrida \cite{derdiff} for random walks in periodic one-dimensional systems with arbitrary fixed transition rates.

Derrida's result.
For completeness, we present below the Derrida's general result for the realization-dependent $D_N$ for random walks evolving in a continuous time on a finite periodic chain with $N$ sites with arbitrary fixed transition rates $W_{x,x\pm1}$.
As shown by Derrida  \cite{derdiff}, the realization-dependent diffusion coefficient in such a setting has the following form 
\begin{equation}
	D_N=\frac{1}{\left( \sum_{n=1}^N r_n \right)^2}
	\left(
	V \sum_{n=1}^N u_n \sum_{i=1}^N i\, r_{n+i}
	+ N \sum_{n=1}^N W_{n,\,n+1}\,u_n\,r_n
	\right)
	- V\,\frac{N+2}{2},
	\label{eq:Dcontinuoustime}
\end{equation}
where  the functions \( u_n \) and \( r_n \) are defined as
\begin{equation}
	u_n=\frac{1}{W_{n,\,n+1}}\left[ 1+\sum_{i=1}^{N-1}
	\prod_{j=1}^i\left(\frac{W_{n-j+1,\,n-j}}{W_{n-j,\,n-j+1}}\right) \right],
	\label{eq:ucontinuoustime}
\end{equation}
\begin{equation}
	r_n=\frac{1}{W_{n,\,n+1}}\left[ 1+\sum_{i=1}^{N-1}
	\prod_{j=1}^i\left(\frac{W_{n+j,\,n+j-1}}{W_{n+j,\,n+j+1}}\right) \right],
	\label{eq:rcontinuoustime}
\end{equation}
while $V$ is the mean (averaged over thermal noise) velocity for a given realization of disorder :
\begin{equation}
	V=\frac{N}{\sum_{n=1}^N r_n}
	\left[ 1 - \prod_{n=1}^N\left(\frac{W_{n+1,\,n}}{W_{n,\,n+1}} \right) \right].
	\label{eq:Vcontinuoustime}
\end{equation}

Model I.
For the Model I the transition rates are defined in Eq. \eqref{rates}.
Using this definition, we find
\begin{equation}
	\prod_{n=1}^N\frac{W_{n+1,\,n}}{W_{n,\,n+1}} = \prod_{n=1}^N \frac{\phi_{n+1}^2}{\phi_{n}^2} = 1,
\end{equation}
which implies that the mean realization-dependent velocity for this model is exactly equal to zero,  as it should.

We proceed to compute the quantities needed for the evaluation of the diffusion coefficient.  The first is \( r_n \), which becomes
\begin{align}
	r_n &= \frac{1}{W_0} \frac{\phi_{n+1}}{\phi_{n}} \left[ 1 + \sum_{i=1}^{N-1} \prod_{j=1}^i \left( \frac{\phi_{n+j}}{\phi_{n+j-1}} \cdot \frac{\phi_{n+j+1}}{\phi_{n+j}} \right) \right] \nonumber\\
	&= \frac{1}{W_0} \frac{\phi_{n+1}}{\phi_{n}} \left[ 1 + \frac{\phi_{n+2}}{\phi_n} + \frac{\phi_{n+2}\,\phi_{n+3}}{\phi_n\,\phi_{n+1}} + \dots + \frac{\phi_{n+N-1}\,\phi_{n+N}}{\phi_n\,\phi_{n+1}} \right] \nonumber\\
	&= \frac{1}{W_0} \frac{1}{\phi_n^2} \left( \phi_n \phi_{n+1} + \phi_{n+1} \phi_{n+2} + \dots + \phi_{n+N-1} \phi_{n+N} \right) = \frac{1}{W_0} \frac{1}{\phi_n^2} H_{0,N},
	\label{eq:rmodelI}
\end{align}
where \( H_{0,N} \) is given in Eq. \eqref{H}.
Further on, from Eq. \eqref{eq:rmodelI} we have that
\begin{equation}
	\sum_{n=1}^N r_n = \frac{1}{W_0} H_{0,N} \sum_{n=1}^N \frac{1}{\phi_n^2} = \frac{1}{W_0} H_{0,N} K_{0,N},
\end{equation}
where we used the shortening
\begin{align}
	K_{n,n+N} &\equiv \sum_{i=0}^{N-1}\frac{1}{\phi^2_{n+i}}=K_{0,N} \,. \label{eq:defK}
	\end{align}
Next, we calculate \( u_n \) by substituting the rates into Eq. \eqref{eq:ucontinuoustime} which gives
\begin{align}
	u_n &= \frac{1}{W_0} \frac{\phi_{n+1}}{\phi_n} \left[ 1 + \sum_{i=1}^{N-1} \prod_{j=1}^i \left( \frac{\phi_{n-j+1}^2}{\phi_{n-j}^2} \right) \right] \nonumber\\
	&= \frac{1}{W_0} \frac{\phi_{n+1}}{\phi_n} \left[ 1 + \frac{\phi_n^2}{\phi_{n-1}^2} + \frac{\phi_n^2 \phi_{n-1}^2}{\phi_{n-1}^2 \phi_{n-2}^2} + \dots + \frac{\phi_n^2 \cdots \phi_{n-N+2}^2}{\phi_{n-1}^2 \cdots \phi_{n-N+1}^2} \right] \nonumber\\
	&= \frac{1}{W_0} \phi_n \phi_{n+1} \sum_{i=0}^{N-1} \frac{1}{\phi_{n-i}^2} = \frac{1}{W_0} \phi_n \phi_{n+1} K_{0,N}.
\end{align}
Lastly, upon some algebra, we find that
\begin{align}
	\sum_{n=1}^N W_{n,n+1} u_n r_n 
	&= \sum_{n=1}^N \frac{\phi_n}{\phi_{n+1}} \cdot \phi_n \phi_{n+1} K_{0,N} \cdot \frac{1}{W_0} \frac{1}{\phi_n^2} H_{0,N} \nonumber\\
	&= \sum_{n=1}^N K_{0,N} \cdot \frac{1}{W_0} H_{0,N} = \frac{N}{W_0} K_{0,N} H_{0,N}.
\end{align}
Combining all the above results, we obtain the following expression for the realization-dependent diffusion coefficient in the Model I:
\begin{align}
	D_N^{(I)} &= \frac{N}{\left( \sum_{n=1}^N r_n \right)^2} \left( \sum_{n=1}^N W_{n,n+1} u_n r_n \right) 
	= \frac{N}{\left( \frac{1}{W_0} H_{0,N} K_{0,N} \right)^2} \cdot \frac{N}{W_0} K_{0,N} H_{0,N}\nonumber \\
	&= \frac{W_0 N^2}{H_{0,N} K_{0,N}} \,,
\end{align}
which is exactly the expression \eqref{DRF} in the main text.

Model II. 
In this dynamical scenario with randomized stepping-times, the transition rates $W_{x,x \pm 1} = p_{x,x \pm 1}/\delta t$, where $p_{x,x \pm 1}$ are the transition probabilities of an auxiliary random process that obey Eqs. \eqref{3} and \eqref{33}.
Firstly, we notice that
\begin{align}
	\prod_{k=1}^N \frac{p_{k,k-1}}{p_{k,k+1}} 
	&= \prod_{k=1}^N 
	\frac{\phi_{k+1}}{\phi_{k-1} + \phi_{k+1}} \cdot
	\frac{\phi_{k-1} + \phi_{k+1}}{\phi_{k-1}} 
	= \prod_{k=1}^N \frac{\phi_{k+1}}{\phi_{k-1}} = 1 \,,
\end{align}
which signifies that the mean realization-dependent velocity is exactly equal to zero, as it should.

Next, we evaluate \(r_n\):
\begin{align}
	r_n &= \frac{1}{p_{n,n+1}} \left( 1 + \sum_{i=1}^{N-1} 
	\prod_{j=1}^i \frac{p_{n+j,n+j-1}}{p_{n+j,n+j+1}} \right) = \frac{\phi_{n-1} + \phi_{n+1}}{\phi_{n-1}} 
	\left( 1 + \sum_{i=1}^{N-1} \prod_{j=1}^i 
	\frac{\phi_{n+j+1}}{\phi_{n+j-1}} \right) \nonumber\\
	&= \frac{\phi_{n-1} + \phi_{n+1}}{\phi_{n-1} \phi_n \phi_{n+1}} H_{n,n+N} = \left( \frac{1}{\phi_n \phi_{n+1}} + \frac{1}{\phi_n \phi_{n-1}} \right) H_{0,N}.
\end{align}
Summing the above expression over \(n\) we find:
\begin{align}
	\sum_{n=1}^N r_n 
	&= \sum_{n=1}^N \left( \frac{1}{\phi_n \phi_{n+1}} + \frac{1}{\phi_n \phi_{n-1}} \right) H_{0,N} = 2H_{0,N} Q_{0,N} \,,
\end{align}
with
\begin{align}
	Q_{n,n+N} &\equiv \sum_{i=0}^{N-1}\frac{1}{\phi_{n+i}\phi_{n+1+i}}=Q_{0,N} \,. \label{eq:defQ}
	\end{align}
Next, for \(u_n\) we find
\begin{align}
	u_n &= \frac{1}{p_{n,n+1}} \left( 1 + \sum_{i=1}^{N-1} 
	\prod_{j=1}^i \frac{p_{n+1-j,n-j}}{p_{n-j,n-j+1}} \right) \nonumber\\
	&= \frac{\phi_{n-1} + \phi_{n+1}}{\phi_{n-1}} \left( 1 + \sum_{i=1}^{N-1} 
	\prod_{j=1}^i \frac{\phi_{n-j+2}}{\phi_{n-j} + \phi_{n-j+2}} 
	\cdot \frac{\phi_{n-j-1} + \phi_{n-j+1}}{\phi_{n-j-1}} \right) \nonumber\\
	&= 2\phi_n \phi_{n+1} Q_{0,N}.
\end{align}
Lastly, we evaluate the sum 
\begin{align}
	\sum_{n=1}^N p_{n,n+1} u_n r_n 
	&= \sum_{n=1}^N \frac{\phi_{n-1}}{\phi_{n-1} + \phi_{n+1}} 
	\cdot 2\phi_n \phi_{n+1} Q_{0,N} 
	\cdot \frac{\phi_{n-1} + \phi_{n+1}}{\phi_{n-1} \phi_n \phi_{n+1}} H_{0,N} \nonumber\\
	&= 2N H_{0,N} Q_{0,N}.
\end{align}
Combining the above expressions, we find the following 
compact result for the realization-dependent  
diffusion coefficient in the Model II :
\begin{align}
	D_N^{(II)} 
	&= \frac{N}{\left( \sum_{n=1}^N r_n \right)^2} 
	\sum_{n=1}^N p_{n,n+1} u_n r_n = \frac{N}{(2H_{0,N} Q_{0,N})^2} \cdot 2N H_{0,N} Q_{0,N} \nonumber\\
	&= \frac{N^2}{2 H_{0,N} Q_{0,N}} \,,
\end{align}
which is our Eq. \eqref{DDD} in the main text.

Model III.
In the Gaussian random trap model the transition rates obey Eq. \eqref{rates3}.
It is straightforward to verify that the product
\begin{equation}
	\prod_{n=1}^N \frac{W_{n+1,n}}{W_{n,n+1}} 
	= \prod_{n=1}^N \frac{\phi_{n+1}^2}{\phi_n^2}= 1 \,,
\end{equation}
which implies that the mean velocity for this model is also equal to zero for any realization of disorder. 

Next, for the quantity \(r_n\) we find
\begin{align}
	r_n 
	&= \frac{1}{W_0} \frac{1}{\phi_n^2} \left[ 1 + \sum_{i=1}^{N-1} \prod_{j=1}^i \left( \frac{\phi_{n+j}^2}{\phi_{n+j}^2} \right) \right] = \frac{1}{W_0} \frac{N}{\phi_n^2} \,,
\end{align}
and consequently, 
\begin{align}
	\sum_{n=1}^N r_n 
	&= \sum_{n=1}^N \frac{1}{W_0} \frac{N}{\phi_n^2} = \frac{N}{W_0} K_{0,N} \,.
\end{align}
Next, we have that \(u_n\) obeys
\begin{align}
	u_n 
	&= \frac{1}{W_0} \frac{1}{\phi_n^2} \left[ 1 + \sum_{i=1}^{N-1} \prod_{j=1}^i \frac{\phi_{n-j+1}^2}{\phi_{n-j}^2} \right] \nonumber\\
	&= \frac{1}{W_0} \frac{1}{\phi_n^2} \left[ 1 + \frac{\phi_n^2}{\phi_{n-1}^2} + \frac{\phi_n^2 \phi_{n-1}^2}{\phi_{n-1}^2 \phi_{n-2}^2} + \dots + \frac{\phi_n^2 \cdots \phi_{n-N+2}^2}{\phi_{n-1}^2 \cdots \phi_{n-N+1}^2} \right] \nonumber\\
	&= \frac{1}{W_0} \sum_{i=0}^{N-1} \frac{1}{\phi_{n-i}^2} = \frac{1}{W_0} K_{0,N}.
\end{align}
Lastly, we get
\begin{align}
	\sum_{n=1}^N W_{n,n+1} u_n r_n 
	&= \sum_{n=1}^N W_0 \phi_n^2 \cdot \frac{1}{W_0} K_{0,N} \cdot \frac{1}{W_0} \frac{N}{\phi_n^2} \nonumber\\
	&= \frac{N^2}{W_0} K_{0,N}.
\end{align}
Combining  all the above expressions we find from Eq.  \eqref{eq:Dcontinuoustime} the following expression for the realization-dependent diffusion coefficient in the Model III:
\begin{align}
	D_N^{(III)} 
	&= \frac{N}{\left(\sum_{n=1}^N r_n \right)^2} \left( \sum_{n=1}^N W_{n,n+1} u_n r_n \right) \nonumber\\
	&= \frac{N W_0^2}{N^2 K_{0,N}^2} \cdot \frac{N^2}{W_0} K_{0,N} = \frac{W_0 N}{K_{0,N}} \,,
\end{align}
which is the expression \eqref{333} in the main text.


\section*{References}

\begin{thebibliography}{99}


\bibitem{as} P. W. Anderson, {\it Absence of diffusion in certain random lattices}, Phys. Rev. {\bf 109}, 1492 (1958).
\bibitem{der} B. Derrida and J. M. Luck, {\it Diffusion on a random lattice: Weak-disorder expansion in arbitrary dimension}, Phys. Rev. B {\bf 28}, 7183 (1983).
\bibitem{luck} J. M. Luck, {\it Diffusion in a random medium: a renormalization group approach}, Nucl. Phys. B {\bf 225},  169 (1983).
\bibitem{derdiff} B. Derrida, {\it Velocity and diffusion constant of a periodic one-dimensional hopping model}, J. Stat. Phys. {\bf 31},  433 (1983).
\bibitem{fis} D.S. Fisher, {\it Random walks in random environments}, Phys. Rev. A {\bf 30}, 960 (1984).
\bibitem{durrett1} R. Durrett, {\it Multidimensional Random Walks
in Random Environments
with Subclassical Limiting Behavior}, Commun. Math. Phys. {\bf 104}, 87  (1986). 
\bibitem{durrett2} M. Bramson and R. Durrett, {\it Random Walk in Random Environment:
A Counterexample?}, Commun. Math. Phys. {\bf 119}, 199 (1988). 
\bibitem{zwan} R. Zwanzig, {\it Diffusion in a rough potential.}, Proc. Natl. Acad. Sci. USA {\bf 85},  2029 (1988).
\bibitem{touya} C. Touya and D. S. Dean, {\it Dynamical transition for a particle in a squared
Gaussian potential}, J. Phys. A: Math. Theor. {\bf 40}, 919 (2007).
\bibitem{dean3} D. S. Dean, I. T. Drummond and R. R. Horgan, {\it Continuum Derrida approach to drift and diffusivity in
random media}, J. Phys. A: Math. Gen. {\bf 30}, 385 (1997).


\bibitem{sinai} Ya. G. Sinai, {\it The Limiting Behavior of a One-Dimensional Random Walk in a Random Medium},  Theor. Prob. Appl. {\bf 27},  256 (1982).
\bibitem{pom} B. Derrida and Y. Pomeau, {\it Classical diffusion on a random chain},  Phys. Rev. Lett. {\bf 48}, 627 (1982).
\bibitem{enzo1} E. Marinari, G. Parisi, D. Ruelle  and P. Windey, {\it Random Walk in a Random Environment and $1/f$
Noise},  Phys. Rev. Lett. {\bf 50},  1223 (1983).
\bibitem{enzo2} E. Marinari, G. Parisi, D. Ruelle and P. Windey, {\it On the Interpretation of $1/f$ Noise}, Commun. Math. Phys. {\bf 89}, 1  (1983).
\bibitem{fish} D. S. Fisher, 
D. Friedan, Z. Qiu, S. J. Shenker and S. H. Shenker, {\it Random walks in two-dimensional random environments with constrained drift forces}, Phys. Rev. A {\bf 31}, 3841 (1985).
\bibitem{kesten} H. Kesten, {\it The limit distribution of Sinai's random walk in random environment}, Physica A {\bf 138},  299 (1986).
\bibitem{pel1} P. Le Doussal, L. F. Cugliandolo and L. Peliti, {\it Dynamics of particles and manifolds in random force fields}, Europhys. Lett. {\bf 39}, 111 (1997).
\bibitem{deandean} A. Comtet and D. S. Dean, {\it Exact results on Sinai's diffusion},
J. Phys. A: Math. Gen. {\bf 31}, 8595 (1998).
\bibitem{d0} G. Oshanin, A. Mogutov and M. Moreau, {\it Steady flux in a continuous-space Sinai chain}, J. Stat. Phys. {\bf 73},  379 (1993).
\bibitem{d1} C. Monthus and A. Comtet, {\it On the flux distribution in a one dimensional disordered system}, J. Phys. I France {\bf 4}, 635 (1994).
\bibitem{d2} G. Oshanin, A. Rosso and G. Schehr, {\it Anomalous Fluctuations of Currents in Sinai-Type Random Chains with Strongly Correlated Disorder}, 
Phys. Rev. Lett. {\bf 110}, 100602 (2013). 
\bibitem{d3} D. S. Dean, S. Gupta, G. Oshanin, A. Rosso and  G. Schehr, {\it Diffusion in periodic, correlated random forcing landscapes},
J. Phys. A: Math. Theor. {\bf 47}, 372001 (2014).


\bibitem{jpb} J. P. Bouchaud, {\it Weak ergodicity breaking and aging in disordered systems}, J. Phys. I France
{\bf 2}, 1705 (1992).
\bibitem{leticia} L. Cugliandolo and P. Le Doussal, {\it Large time off-equilibrium dynamics of a particle diffusing in a random potential},  Phys. Rev. E {\bf 53}, 1525 (1996).
\bibitem{mont} C. Monthus and J.-P. Bouchaud, {\it Models of traps and glass phenomenology},
J. Phys. A: Math. Gen. 29 3847 (1996).
\bibitem{bertin} E. Bertin and J-P. Bouchaud, {\it Dynamical ultrametricity in the critical trap model},
J. Phys. A: Math. Gen. 35 3039 (2002).
\bibitem{bertin2} E. M. Bertin and J-P. Bouchaud, {\it Sub-diffusion and localization in the one-dimensional trap model}, 
Phys. Rev. E {\bf 67}, 026128 (2003).
\bibitem{ralf10} H. Kr\"usemann, A. Godec and R. Metzler, {\it First-passage statistics for aging diffusion in systems with annealed and quenched disorder}, Phys. Rev. E {\bf 89}, 040101(R) (2014).
\bibitem{ralf11}  H. Kr\"usemann, A. Godec and R. Metzler, {\it Ageing first passage time density in
continuous time random walks and
quenched energy landscapes}, J. Phys.  A : Math. Theor. {\bf 48},   285001 (2015). 
\bibitem{ralf12} A. Godec and R. Metzler,   {\it Optimization and universality of Brownian search in a basic model of quenched
heterogeneous media},  Phys. Rev. E {\bf 9}1, 052134 (2015).
\bibitem{aki1} T. Akimoto, E. Barkai, and K. Saito, {\it Universal Fluctuations of Single-Particle Diffusivity in Quenched Environment}, Phys. Rev. Lett. {\bf 117}, 180602 (2016).
\bibitem{aki2} T. Akimoto, E. Barkai, and K. Saito, {\it Non-Self Averaging and Ergodicity in Quenched Trap Model with Finite System Size}, Phys. Rev. E {\bf 97}, 052143 (2018).
\bibitem{ralf13} S. Park, X. Durang, R. Metzler and J.-H. Jeon, {\it Fickian yet non-Gaussian diffusion in an annealed heterogeneous environment};  	arXiv:2503.15366 


\bibitem{mdm} G. Matheron and G. de Marsily, {\it Is transport in porous medium always diffusive: 
A counterexample}, Water Resources Research {\bf 16}, 901 (1980). 
\bibitem{blu}  G. Oshanin and A. Blumen, {\it Rouse chain dynamics in layered random flows}, Phys. Rev. E {\bf 49},  
4185 (1994).
\bibitem{kay} K. J. Wiese and P. Le Doussal, {\it Polymers and manifolds in static random flows:
a renormalization group study},  Nucl. Phys.  B {\bf 552}, 529 (1999). 

\bibitem{man1} J.-P. Bouchaud, A. Georges, J. Koplik, A. Provata and S. Redner, {\it Superdiffusion in random velocity fields},  Phys. Rev. Lett. {\bf 64},  2503 (1990).
\bibitem{man2} C. Mejia-Monasterio, S. Nechaev, G. Oshanin and O. Vasilyev, {\it Tracer diffusion on a crowded random Manhattan lattice},
New J. Phys. {\bf 22}, 033024 (2020).

\bibitem{r1} Y. Kafri, D. K. Lubensky and 
D. R. Nelson, {\it Dynamics of Molecular Motors and Polymer Translocation with
Sequence Heterogeneity},  Biophys. J. {\bf 
86}, 3373  (2004).
\bibitem{r2} Y. Kafri and D. R. Nelson, {\it Sequence heterogeneity and the dynamics of
molecular motors},  J. Phys.: Condens. Matter {\bf 17},  S3871 (2005).
\bibitem{r3}  Y. Kafri, D. K.  Lubensky and D. R. Nelson, {\it Dynamics of molecular motors with finite processivity on heterogeneous tracks}, Phys. Rev. E  {\bf 71},  041906 (2005).
\bibitem{r4} M. Slutsky, M. Kardar and L. A. Mirny, {Diffusion in correlated random potentials, with applications to DNA}, 
Phys. Rev.  E {\bf 69}, 061903 (2004).
\bibitem{22} T. Hwa, E. Marinari, K. Sneppen and L.-h. Tang, Localization of denaturation bubbles in random
DNA sequences, Proc. Natl. Acad. Sci. USA 100, 4411 (2003).
\bibitem{pdg} P. G. de Gennes, Brownian motion of a classical particle through potential barriers. Application
to the helix-coil transitions of heteropolymers, J. Stat. Phys. 12, 463 (1975).
\bibitem{d11} G. Oshanin and S. Redner, {\it Helix or coil? Fate of a melting heteropolymer}, Eur. Phys. Lett. {\bf 85}, 10008 (2009).



\bibitem{al} S. Alexander, J. Bernasconi, W.R. Schneider and R. Orbach, {\it Excitation dynamics in random one-dimensional systems},  Rev.  Mod.  Phys. {\bf 53}, 175 (1981).
\bibitem{pel} L. Peliti, {\it Self-avoiding walks}, Phys. Reports  {\bf 103}, 225 (1984).
\bibitem{kehr} J. W. Haus and K. W. Kehr, {\it Diffusion in regular and disordered lattices}, Phys. Reports {\bf 150}, 263 (1987).
\bibitem{bou} J.-P. Bouchaud and A. Georges, {\it Anomalous diffusion in disordered media: Statistical mechanisms, models and physical applications}, Phys. Reports {\bf 195}, 127 (1990).
\bibitem{hughes} B. D. Hughes, {\it Random Walks and Random Environments}, Vol. 1 (Oxford Science Publ., Clarendon Press, Oxford, 1995).
\bibitem{dean} D. S. Dean, I. T. Drummond and R. R. Horgan, {\it Effective transport properties for
diffusion in random media}, J. Stat. Mech. (2007) P07013; doi:10.1088/1742-5468/2007/07/P07013
\bibitem{bogachev} L. V. Bogachev, {\it Random Walks in Random Environments}, Encyclopedia of Mathematical Physics,  J.-P. Francoise, G. Naber, and S. T. Tsou, Eds., 
{\bf 4}, (Elsevier, Oxford, 2006), p. 353.

\bibitem{baruch} A. Valov, N. Levi and B. Meerson, {\it Thermally activated particle motion in biased correlated Gaussian disorder potentials}, Phys. Rev. E {\bf 110}, 024138 (2024).

\bibitem{olson} C. Reichhardt and C. J. Olson Reichhardt, {\it Active matter transport and jamming on disordered landscapes},  Phys. Rev. E {\bf 90}, 012701 (2014).
\bibitem{licata} N. Licata, B. Mohari, C. Fuqua, and S. Setayeshgar, {\it Diffusion of Bacterial Cells in Porous Media}, Biophys. J. {\bf 110}, 247 (2016). 
\bibitem{zeitz} M. Zeitz, K. Wolff, and H. Stark, {\it Active Brownian particles moving in a random Lorentz gas}, European Phys. J. E {\bf 40}, 23 (2017). 
\bibitem{sant} D. Santillan, {\it Dispersion of run-and-tumble microswimmers through disordered media}, Phys. Rev. E {\bf 108}, 064608 (2023).
\bibitem{mat} H. H. Mattingly, {\it Coarse-graining bacterial diffusion in disordered media to surface states}, Proc. Natl. Acad. Sci. USA {\bf 122} (12), e2407313122 (2025).

\bibitem{kalaj} S. Kalaj, E. Marinari, G. Oshanin  and L. Peliti, {\it A simple model of a sequence-reading diffusion: non-self-averaging and self-averaging properties}, 
New J. Phys.  {\bf 27}, 043022 (2025).

\bibitem{1} R. Zwanzig,  
{\it Diffusion past an entropy barrier}, J. Phys. Chem. {\bf 96}, 3926 (1992).

\bibitem{2} M. Bauer, A. Godec and R. Metzler, {\it Diffusion of finite-size particles in two-dimensional channels with random wall configurations}, Phys. Chem. Chem. Phys. {\bf  16}, 6118 (2014).


\bibitem{mw} E. W. Montroll and  G. H. Weiss, {\it Random Walks on Lattices. II}, J. Math. Phys. {\bf 6},  167 (1965).

\bibitem{katja} D. Bedeaux, K. Lakatos-Lindenberg and K. E. Shuler, {\it On the Relation between Master Equations and Random Walks and Their Solutions}, J. Math. Phys. {\bf 12}, 2116 (1971).



\bibitem{pastur} I. M. Lifshits, S. A. Gredeskul and L. A. Pastur, {\it Introduction to the Theory of Disordered Systems},
(Wiley-VCH, New York, 1988).

\bibitem{katja2}  K. Lindenberg and B. J. West, {\it The First, The Biggest, and Other Such Considerations}, J. Stat. Phys. {\bf 42}, 201 (1986).
\bibitem{redner} S. Redner, {\it A Guide to First Passage Processes}, (Cambridge University Press, Cambridge, 2001).
\bibitem{metz}  R. Metzler, G. Oshanin and S. Redner, eds., {\it First-Passage Phenomena and Their Applications},
(Singapore, World Scientific, 2014)
\bibitem{olivier} O. B\'enichou and R. Voituriez, {\it From first-passage times of random walks in confinement to geometry-controlled kinetics}, Phys. Reports {\bf 539}, 225 (2014). 
\bibitem{denis} D. Grebenkov, R. Metzler and G. Oshanin, eds., {\it Target Search Problems}, (Springer, Cham, 2024).
\bibitem{weiss} G. H. Weiss, {\it First Passage Time Problems in Chemical Physics}, Adv. Chem. Phys. {\bf 13}, 1 (1966).


\bibitem{LJ} S. Lifson and J. L. Jackson, {\it On the self-diffusion of ions in a polyelectrolyte solution}, J. Chem.
Phys. {\bf 36}, 2410 (1962).
\bibitem{festa} R. Festa and E. Galleani d'Agliano, {\it  Diffusion coefficient for a Brownian particle  in a periodic field of force}, Physica A {\bf 90}, 229 (1978).


\bibitem{sasha} J. R. Kalnin and A. M. Berezhkovskii, {\it On the relation between Lifson-Jackson and Derrida formulas for effective diffusion coefficient}, J. Chem. Phys. {\bf 139}, 196101 (2013).

\bibitem{gil} D. T. Gillespie, Daniel T. (2007). {\it Stochastic Simulation of Chemical Kinetics}, Ann. Rev. Phys. Chem. {\bf 58},  35 (2007).

\bibitem{7}  C. Mejia-Monasterio, G. Oshanin and G. Schehr, {\it First passages for a search by a swarm of independent random searchers}, 
J. Stat. Mech.: Theor. Exp. {\bf 2011} (06), P06022.

\bibitem{8} T. G. Mattos, C. Mejia-Monasterio, R. Metzler  and G. Oshanin, {\it First passages in bounded domains: When is the mean first passage time meaningful?},
Phys. Rev. E {\bf  86},  031143 (2012).

\bibitem{vanKampen1992}
N. G. van Kampen, {\it Stochastic Processes in Physics and Chemistry},
(North-Holland Personal Library, Elsevier, Amsterdam, 2007).




\end{thebibliography}
\end{document}